\documentclass[10pt,journal,compsoc]{IEEEtran}
\ifCLASSOPTIONcompsoc
  \usepackage[nocompress]{cite}
\else
  \usepackage{cite}
\fi
\ifCLASSINFOpdf
   \usepackage[pdftex]{graphicx}
\else
   \usepackage[dvips]{graphicx}
\fi
\usepackage{amsmath}
\usepackage{array}

\usepackage{xspace}
\def\etal{et al.\xspace}

\usepackage{amssymb}

\newcommand{\msl}{\mathbf \sigma}

\newcommand{\mslpnew}{\mathbf \sigma_{\tilde{p}}}
\newcommand{\mv}{\mathbf v}

\newcommand{\mxpnew}{\mathbf x_{\tilde{p}}}

\newcommand{\mxpinew}{\mathbf x_{\tilde{p}i} }

\newcommand{\mn}{\mathbf u }

\newcommand{\mwnew}{ w_j(\mxpinew) } 

\newcommand{\mpcsnew}{P(C = c_j | S = \tilde{p}_i)}
\newcommand{\R}{\mathbb{R}}

\usepackage[hang,tight,sf,SF,scriptsize]{subfigure}
\newlength{\lengthgoodgap}
\newlength{\lengthlittlegap}
\usepackage{booktabs}

\usepackage{url}

\usepackage[dvipsnames]{xcolor}
\usepackage{ulem}

\usepackage{listings}
\definecolor{codegreen}{rgb}{0,0.6,0}
\definecolor{codegray}{rgb}{0.5,0.5,0.5}
\definecolor{codepurple}{rgb}{0.58,0,0.82}
\definecolor{backcolour}{rgb}{0.95,0.95,0.92}

\lstdefinestyle{mystyle}{
    backgroundcolor=\color{backcolour},   
    commentstyle=\color{codegreen},
    keywordstyle=\color{magenta},
    numberstyle=\tiny\color{codegray},
    stringstyle=\color{codepurple},
    basicstyle=\ttfamily\footnotesize,
    breakatwhitespace=false,         
    breaklines=true,                 
    captionpos=b,                    
    keepspaces=true,                 
    numbers=left,                    
    numbersep=5pt,                  
    showspaces=false,                
    showstringspaces=false,
    showtabs=false,                  
    tabsize=2
}
 
\begin{document}
%
\title{A Discrete Probabilistic Approach to Dense Flow Visualization}

\author{Daniel~Preu\ss{}, Tino~Weinkauf, and Jens~Kr\"{u}ger,~\IEEEmembership{Member,~IEEE,}
\IEEEcompsocitemizethanks{\IEEEcompsocthanksitem D. Preu\ss{} and J. Kr\"{u}ger are with COVIDAG, University of Duisburg-Essen, Germany,
Duisburg, 47057.\protect\\
E-mail: daniel.preuss@uni-due.de\protect\\
\hspace*{9mm} jens.krueger@uni-due.de
\IEEEcompsocthanksitem T. Weinkauf is with KTH Royal Institute of Technology, Sweden, Stockholm, 114 28. \protect\\
E-mail: weinkauf@kth.se}
\thanks{Manuscript received September 4, 2019; revised August 26, 2015.}}

%
%

\markboth{IEEE Transactions on Visualization and Computer Graphics,~Vol.~14, No.~8, August~2015}%
{Preuß \MakeLowercase{\textit{et al.}}: A Discrete Probabilistic Approach to Dense Flow Visualization}
%



\IEEEtitleabstractindextext{%
\abstract{%

Dense flow visualization 
is a popular visualization paradigm.
Traditionally, the various models and methods in this area 
use a continuous formulation,
resting upon the solid foundation of functional analysis.
In this work,
we examine a discrete formulation of dense flow visualization.
From probability theory, we derive a similarity matrix that measures the similarity between different points in the flow domain, leading to the discovery of a whole new class of visualization models.
Using this matrix, we propose a novel visualization approach consisting of the computation of spectral embeddings,
i.e., characteristic domain maps,
defined by particle mixture probabilities.
These embeddings are scalar fields that give insight 
into the mixing processes of the flow on different scales. The approach of spectral embeddings is already
well studied in image segmentation, and we see that spectral embeddings are connected to Fourier expansions and frequencies.
We showcase the utility of our method using different 2D and 3D flows.%
}

}

\maketitle

\IEEEdisplaynontitleabstractindextext

%
\IEEEpeerreviewmaketitle

\IEEEraisesectionheading{\section{Introduction}\label{sec:introduction}}

%
%
%
%
\IEEEPARstart{D}{ense}, or texture-based flow visualization (DFV), and particularly
    the Line Integral Convolution (LIC) method, spot noise, and "image-based" flow visualization \cite{Cabral:1993:IVFU, stalling95,Wijk:1991:SNTS,vanWijk:imagebasedflowvis}
    have been proved successful in many scientific and engineering applications.     
    Its wide popularity is the result of such features as 
    suitability for efficient parallel implementation on graphics hardware
    and the possibility to use adaptive resolution.         
    
    Interestingly, despite a vast body of research on the subject
    \cite{Laramee:2004:TSOA} and the fact that DFV is tightly related 
    to the long-established branches of mathematics such as numerical methods,
    to the best of our knowledge, a consistent theoretical framework
    that would allow systematic interpretation and exploration of  different modifications has not yet been proposed.    
    For instance, the net effect of the numerous ingredients, such 
    as the noise interpolation scheme, the kernel shape, and streamline integration 
    sampling methods, on the output image cannot always be predicted.

    The variety of models and methods in this area
    is typically formulated within a continuous setting, whereas
    the gap between the digital nature of the computational world
    and these continuous models is typically bridged by numerical discretization.
    
    In this work, we propose a probabilistic model using conditional expectation computation for the position of a particle on its trajectory and derive a discrete linear algebra formulation. By multiplying this matrix with a noise input vector, we achieve results similar to LIC.
    We demonstrate that this formulation is compelling on its own, and closer examination
    can lead to fruitful insights, connections to image processing,
    and new visualization algorithms.     
    We envision our purely discrete algebraic interpretation
    as a step toward a systematic theoretical framework for DFV.

    This change of paradigm allows for further development.
    We apply probability modeling to discrete images
    to establish a probabilistic relationship between image pixels 
    based on the trajectories of particles seeded in the flow in the cells, 
    corresponding to pixels.
    Then, we explore the visualization images, constructed 
    as an optimal solution, by minimizing the expected difference in the color space
    for cells with similar flow behavior. The optimal solution contains one scalar value for each cell in our image, which is then mapped to colors with a transfer function.

    The obtained images visualize flow mixture patterns of particles
    and can be formally described as the eigenvectors
    of the Laplacian of the particle mixture probability matrix. We then see that the eigenvalues
    of the Laplacian matrix correspond to frequencies, and the eigenvectors are a discretization of the terms of a Fourier expansion.
    These eigenvectors,  widely referred to as spectral embeddings,  
    are a powerful tool 
    for the analysis of different types of graphs. 
    In particular, numerous variations of spectral clustering \cite{Brand:2003:UTSEC},
    due to its success in computer vision in recent years \cite{Luxburg:2007:TSC}, 
    have gained popularity in image segmentation. 

    We find the achieved results interesting for applications 
    and encouraging for further research.      
    Briefly, the two leading contributions of this work are:
    \begin{itemize}
      \item a matrix formulation as a new DFV method,
      \item a novel discrete probabilistic modeling framework 
      for the development and analysis of dense visualization methods.
    \end{itemize}

    \begin{figure*}[t]
        \centering
      \includegraphics[width=\textwidth]{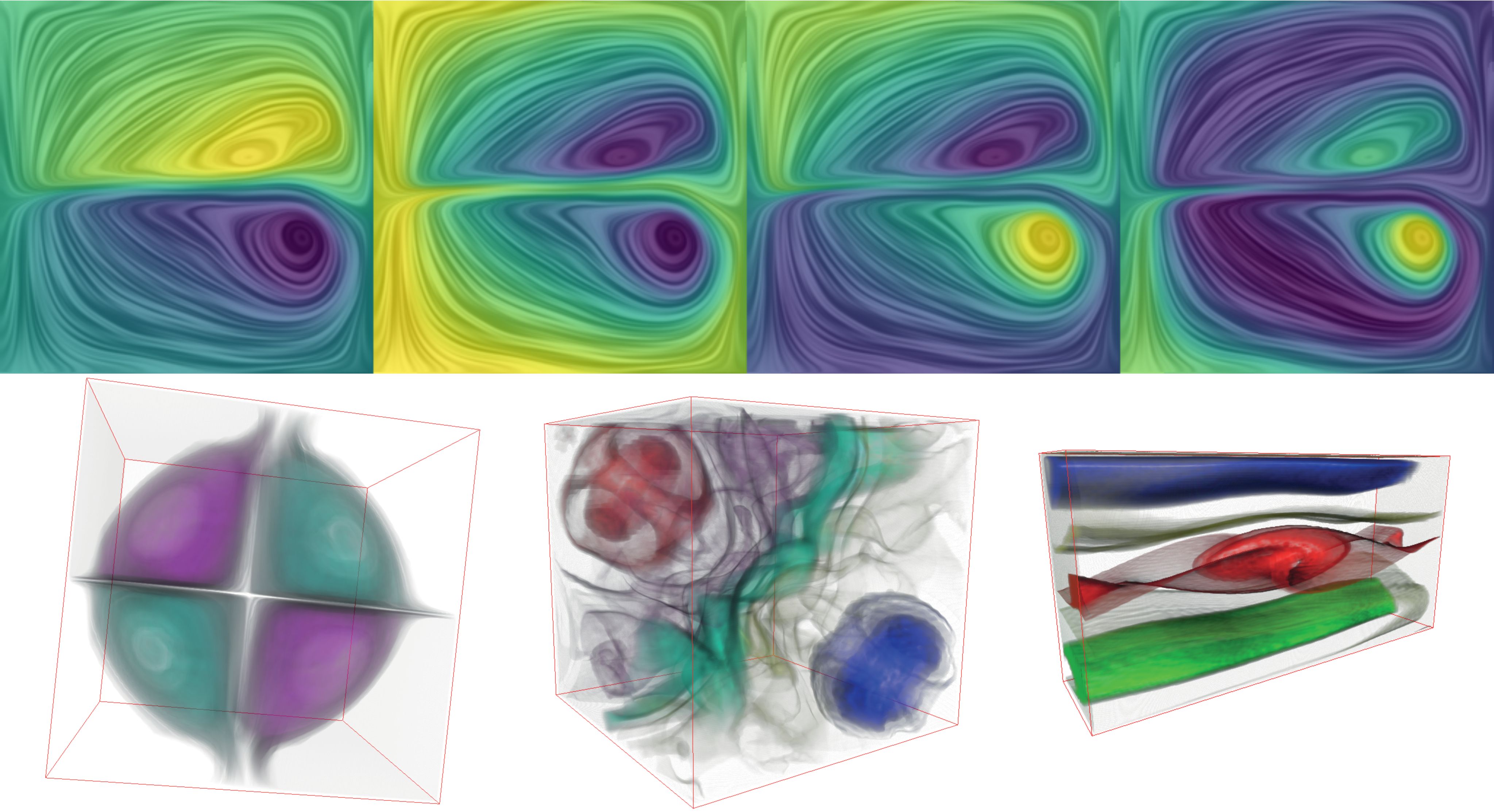}
    \caption{
      The first four eigenvectors of the Laplacian of the mixture probability matrix $H$
      represented as 2D arrays and with the viridis color map applied are shown in the top row and are combined with the output of our probability matrix multiplied by a noise input. 
      The eigenvectors are ordered from left to right by the corresponding eigenvalue.
      In the bottom row, a visualization of different flows using direct volume rendering of the flow graph embeddings is shown.
      The datasets from left to right are 
      flow in a spherical drop, Borromean magnetic field, and the Stuart vortex.
      The embedding volumes were computed at 128 $\times$ 128 $\times$ 128 resolution with a Gaussian kernel of 
      half-length 80 (voxels).
    }
      \label{fig:teaser}
    \end{figure*}

    \begin{figure}[t]
    \centering
    \includegraphics[width=\columnwidth]{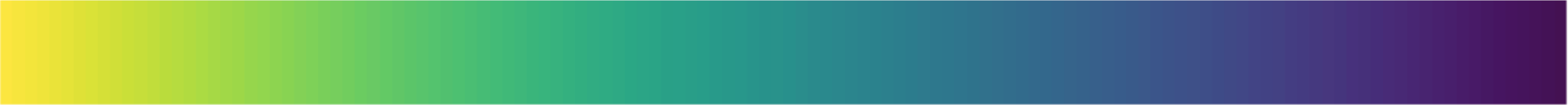}
    \caption{The color bar corresponding to the Viridis colormap.}
    \label{fig:colorbar}
    \end{figure}
    
    The remainder of this paper is structured as follows: 
    In the next section, we provide a brief review of the previous work 
    in the context of dense visualization as well
    as relevant flow and image analysis methods.    
    Then, in Section~\ref{sec:flowvisemb},
    we suggest a novel discrete probabilistic model for flow representation, with applications to dense flow visualization. We derive a matrix from this probabilistic model, which results in similar outputs to LIC and explore its properties, including maximum value preservation. Furthermore, we use the probability that two particles meet in a given cell and the spectral embedding method to correlate different cells in the flow.
    In Section~\ref{sec:implementation}, we discuss the technical details of the method.  
    Finally, we demonstrate the visualization results     
    and discuss future work.
    
    The presented results use the Viridis colormap, seen in \figurename~\ref{fig:colorbar}.
    
\section{Related Work}
  The dense flow visualization paradigm,
  starting with the introduction of spot noise~\cite{Wijk:1991:SNTS} and 
  LIC~\cite{Cabral:1993:IVFU},
  has undergone significant development over almost 25 years.
  The state-of-the-art report by Laramee \etal \cite{Laramee:2004:TSOA}
  enumerates a broad set of methods derived from these two approaches.
  Mainly, the efforts have been focused on an extension to further dimensions,
  such as 3D LIC \cite{Rezk-Salama:1999:IEOV}, LIC on surfaces \cite{Forssell:1994:VFCG}, LIC on time-varying flow fields \cite{UFLIC}, and efficient implementation \cite{Stalling:1995:FARI, Weiskopf:2006:GPUB}.
  
  Although new competitive methods have appeared, 
  LIC in its various modifications has remained a workhorse of DFV.
  Several works concerning its theoretical grounds and improvement have been published.
  A thorough analysis of the influence of LIC parameters 
  from the signal processing standpoint
  was done by Stalling \cite{Stalling:1998:FTBBA}.
  Okada and Lane \cite{Okada:1997:ELIC} introduced the concept of twofold LIC for image enhancement.
  Its value and benefits for computation acceleration 
  were later recognized by Weiskopf \cite{Weiskopf:2009:ITLI} 
  and Hlawatsch \etal~\cite{Hlawatsch:2010:HLI}.

  Despite the substantial body of knowledge on the topic, 
  there is still room for new models and interpretations.
  One attempt to deal with the lack 
  of a complete theoretical framework for LIC, for example,
  is the empirical quantitative analysis of LIC images \cite{Matvienko:2013:AMFT}.
   
  Based on the generalization of existing DFV techniques, 
  we formulate a discrete probabilistic model of flow mixture in the domain.
  Notably, the efforts to employ probabilistic methods for the analysis of streamline separation
  were previously successful in the work of Reich and Scheuermann \cite{Reich:2012:ASSI}.
  They visualize a measure of convergence and divergence 
  between particles seeded in the neighboring cells (pixels),
  after a number of iterations of a Markov chain 
  over each particle's initial probability distribution. 
  Our matrix model differs from this setting in the following two main aspects: 
  \begin{itemize} 
  \item It embraces the information about the whole integral curve, 
  instead of a particle movement in a single time step;
  \item The time-consuming iterative eigenvector 
  computation process is required only once, but not for every domain cell.
  \end{itemize}
  
  As a result of much lower computational complexity, our method is applicable in 3D.
  From the visualization perspective,
  we propose a global map of the domain with a progressive level of detail 
  characterizing the mixture relationship between the cells in the domain.
  Our technique aims  
  to highlight structures in the underlying data 
  and to provide their visual representation 
  at different scales.     
  Within the existing classification, our visualization approach can be described as partition-based.
  The state-of-the-art report on this topic \cite{Salzbrunn:2008:STPT} 
  names two main subclasses in this area: 
  based on vector value clustering and relying on integral line analysis.
  Our method combines the features of both approaches:
  distinguishing the regions of the flow domain and    
  using the information about particle trajectories
  instead of the vector value. 
  
  The principal idea of flow simplification
  is extensively exploited from a different perspective in topological methods.
  Salzbrunn \etal \cite{Salzbrunn:2008:TSOA} give an outlook on this research, 
  and we name only a few notable results.
  Helman and Hesselink \cite{helman89}
  extracted the critical points and separatrices of 2D vector fields,
  which provided a segmentation into sectors of different flow behaviors.
  Topological simplification \cite{tricoche01a, weinkauf05a}
  is a way to identify the more salient topological features in a flow.
  Recent developments include
  combinatorial vector field topology \cite{reininghaus11a}
  and streamline clustering using Morse connection graphs \cite{Szymczak:2013:VMCG, efficientMorse}
  or streamline predicates \cite{Salzbrunn:2006:SP}. Peng \etal \cite{Peng} cluster flow structures on a surface mesh.
  
  Other approaches simplify or segment the flow
  without referring to topology.
  For example, R\"ossl \etal \cite{Roessl:2012:TVCG}
  group streamlines using their projection into Euclidean space 
  with a Hausdorff distance.
  The methods of Garcke \etal \cite{Garcke:2002:CCMVF},
  Heckel \etal \cite{heckel99},
  and Telea and van Wijk \cite{telea99}
  cluster cells with similar flow vectors. Padberg-Gehle \etal \cite{SegmentationPDEs} extract coherent   
  sets using discretized transfer operators. Diewald \etal \cite{AnisotropicSegmentationPDE} use anisotropic nonlinear 
  diffusion to create similar results to LIC and cluster flow fields in coherent structures.
  Park \etal \cite{Park:2006:SADV} suggested a DFV approach to accentuating flow structures.

  Many approaches use probabilistic models to visualize flow behavior. Höllt \etal \cite{probHollt} visualize the paths a particle, originating in a specific cell, can travel in subsequent steps and the corresponding probabilities. Guo \etal \cite{Guo} use an adaptive and decoupled scheme to accelerate the calculation of stochastic flow maps with Monte Carlo runs. Otto \etal \cite{Otto} visualize uncertain areas in uncertain vector fields, where a cell can have multiple flow vectors, each with its probability.
  
  The technical side of our approach 
  is inspired by the wave interference method in DFV \cite{Matvienko:2012:DFVU},  
  relying on the similar mathematical apparatus of sparse matrix computations.  
  The eigenvector computation, resulting from the analysis of our probabilistic model,
  has a direct correspondence to the spectral embeddings technique
  widely used in the image processing domain. 
  There it consists in the representation of the 
  image segmentation as a graph cut problem,
  with its consequent relaxation using 
  spectral graph theory.
  For instance, the normalized cut method in image segmentation 
  gained wide popularity after the presentation of the approach by Shi and Malik \cite{Shi:1997:NCIS}. 
  Their method works similarly to the here presented algorithm, by first creating a similarity graph, then building a Laplacian matrix from that graph, and finally calculating the eigenvectors. However, they use a different eigenvalue problem by applying matrix $D$ to the right side and process the results differently.
  For introductory reading on the subject, we suggest the tutorial by  Luxburg \cite{Luxburg:2007:TSC}.
  The original graph formulation was followed by  the random walk interpretation 
  of Meila and Shi \cite{Meila:2001:LSRW}, 
  which assigns intuitive meaning to the spectral embeddings,
  relating them to the concept of low conductivity sets.  
  Moreover, Shuman \etal \cite{LaplacianEigenvalueFrequency} relate the Laplacian matrix derived in 
  spectral graph theory to the Fourier transformation and the eigenvalues to frequencies.
  Finally, the theoretical aspects of the algorithm were 
  carefully treated by Brand and Huang \cite{Brand:2003:UTSEC}.  

\begin{figure}[t!]
\centering
\includegraphics[width=\columnwidth]{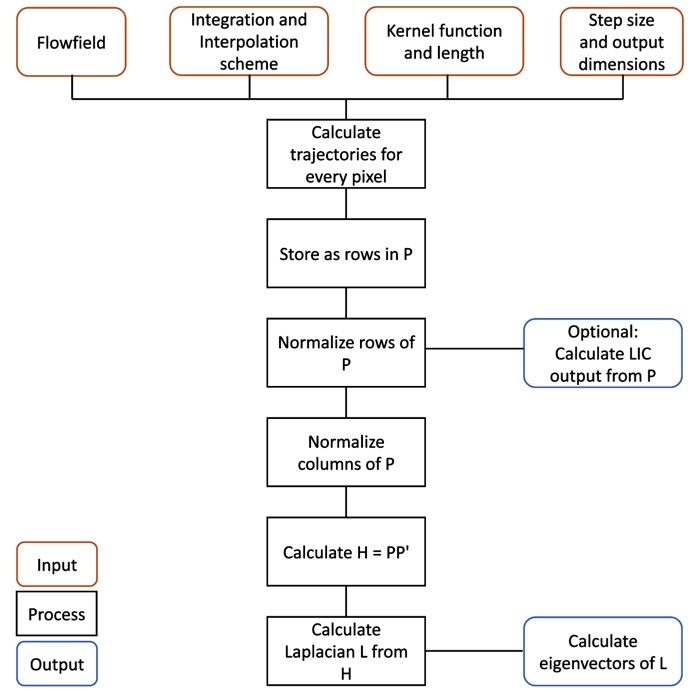}
\caption{The overall structure of the presented algorithm.}
\label{fig:overview}
\end{figure}

\section{Probabilistic Approach to Flow Visualization}\label{sec:flowvisemb}
In this section, we present our contribution to the toolbox of methods for dense flow visualization.
Furthermore, we use our probabilistic model to analyze the discrete flow domain in terms of particle transport (mixture probability). In \figurename~\ref{fig:overview} an overview of the presented algorithm is shown.

\subsection {Probabilistic Model} \label{probModel}
Considering a 2D stationary flow domain,
we introduce a rectangular grid over the domain, 
consisting of cells corresponding to image pixels. Let $N$ be the number of cells in our grid.
We associate a particle $\tilde{p}_i$ with each cell $c_i$
and observe it for a specific interval $L$ before and after
it is registered in some chosen position within the cell.

\begin{figure}[t]   \centering
\includegraphics[width=\columnwidth]{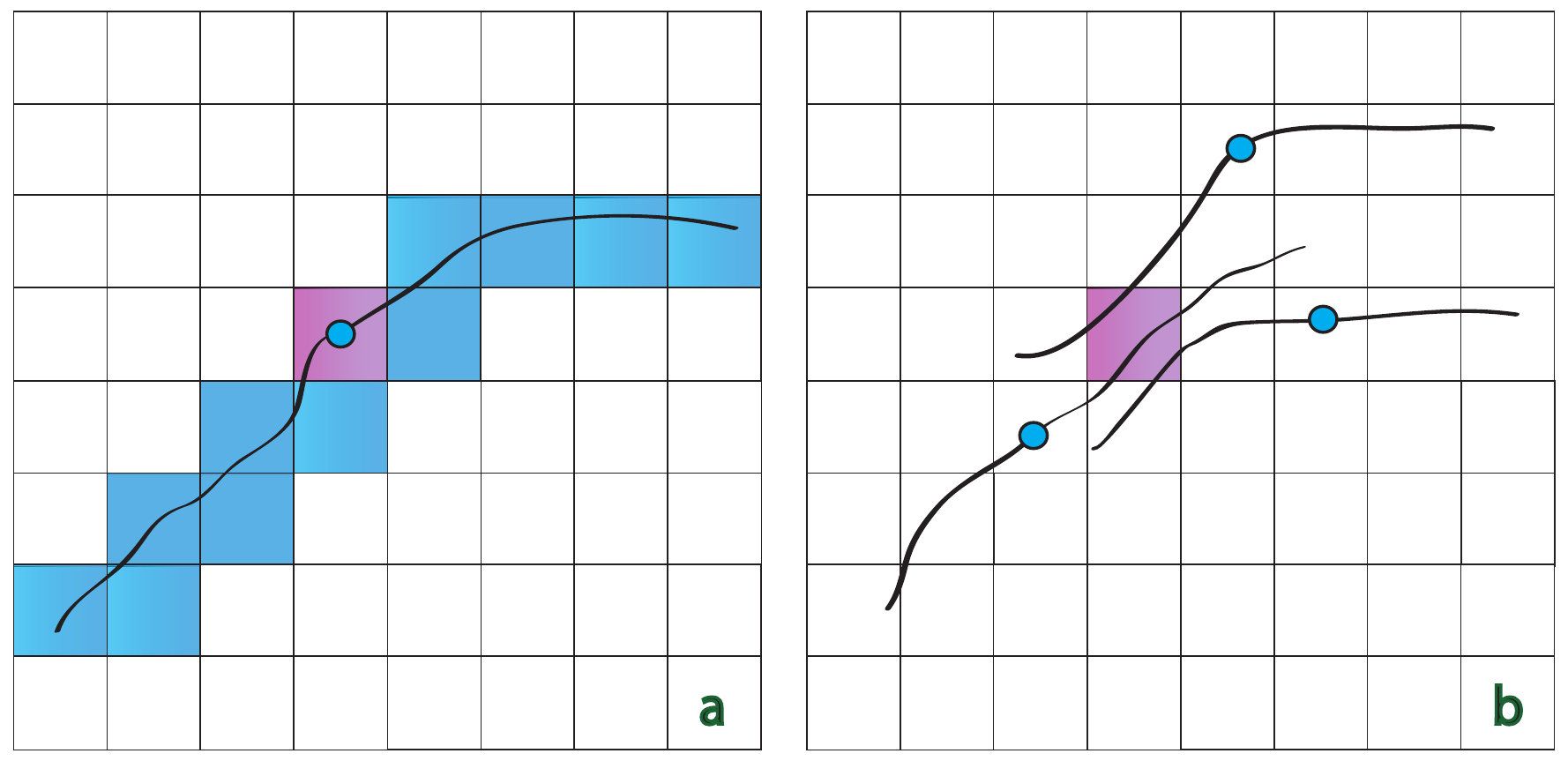} %
\caption{%
a) The blue cells on the trace of a particle 
that contribute to the computation of the image value in the red cell,
where the particle is observed.%
\; b) Several blue particles, each seeded in its own cell, contribute to the value of one red cell.}%
\label{fig:particles}%
\end{figure}

We introduce two random variables: 
$S$, which takes the value on the set of observed particles $\tilde{p}_1,...,\tilde{p}_N$,
and $C$, which takes the value on the set of cells $c_1,...,c_N$.
The particle can visit several other cells
while moving along the streamline, as illustrated by \figurename~\ref{fig:particles}a. The streamline computation contains degrees of freedom in choosing the step size, kernel length $L$, and kernel function.
The kernel length $L$ is defined as an integer, and together with the step size, it defines the distance we observe the particle position on its trajectory.
A kernel function is a measure that defines how probable a particle position is with increasing distance from its origin.

Two common examples are the Box and the Gaussian kernel.
The Box kernel, corresponding to the uniform distribution, 
suggests that each cell on the particle trajectory likely equally captures the position of the particle.
The Gaussian kernel, on the other hand, expresses an increasing uncertainty 
about the particle position, the further away it is from the initial point.

For each cell on the particle trajectory,
we assume a conditional probability $\mpcsnew$,
denoting the probability of a given particle $\tilde{p}_i$ visiting a particular cell $c_j$.
This distribution is expressed by the shape of the chosen kernel and represents 
the measure of relevance of $c_j$ to the trace of $\tilde{p}_i$.

From this point of view, given a discrete 2D signal $U$ defined on the cells $c_i$,
we can compute for each of 
the cells $c_i$, and the particles $\tilde{p}_i$ associated with them,
the conditional expectation
\begin{equation} \label{equ:condExpec}
E_{C|S}[U] =  \sum_j \mpcsnew U(c_j) 
\end{equation}
of the input distribution over a set of cells visited by the particle $\tilde{p}_i$.
As a result, the output values are correlated for two particles
if they produce overlapping traces,
that is, traces sharing mutual cells, similar to the output of LIC.
In particular, the input signal $U$ can represent
the initial position probability distribution of a single particle $\tilde{p}$ over all domain cells,
as, for instance, in the model suggested for streamline separation by Reich and Scheuermann~\cite{Reich:2012:ASSI}.
In this case, the expected value in cell $c_i$  corresponds 
to the probability of particle $\tilde{p}$ arriving in cell $c_i$.
This case is intuitively illustrated with oriented LIC~\cite{wegenkittl97},
considered within the probabilistic framework.
If the sparse input texture represents the distribution of 
possible initial positions of a single particle
and the LIC kernel is asymmetric (backward flow direction only),
 for each output image cell, one possible trajectory history is sampled. The resulting image represents the probability 
that the initial particle arrives at this cell.

In the next section, we show how to compute the conditional probabilities for our probabilistic model and derive a matrix from them.

\subsection{Probability Matrix}

Given a vector field $ \mv$ defined by a map $\mv: \mathbb{R}^2 \rightarrow\mathbb{R}^2$,
its streamlines $\sigma$ can be defined using the arc-length parameterization by 
\begin{align} \label{equ:streamline}
  \frac{d}{ds} \msl(s) = \frac{\mv}{|\mv| } 
\end{align}
\noindent in the regions where $|\mv| \ne 0$.
Throughout the text, the boldface letters denote vectors in $\mathbb{R}^n$ for various $n$.

Here we assume that for each $\mxpnew \in \R^2$ of the image domain,
a streamline $\mslpnew$ can be computed analytically or numerically,
satisfying the initial condition $\mslpnew(0) = \mxpnew$.
The conditional expectation $e_{\tilde{p}}$ of Equation~\ref{equ:condExpec}
is then defined as 
\begin{align}
e_{\tilde{p}}  = \int_{-L}^{L} U(\mslpnew(t))k(t) dt \label{equ:intlic}
\end{align}
where $U(\mathbf x)$ is the input signal, $L$ the chosen kernel length, and $k(t)$ the weighting kernel.

Under mild smoothness assumptions (local Lipschitz condition \cite{Stalling:1995:FARI})
about the left-hand side of Equation~\ref{equ:streamline},
the computed curves are unique, given the initial conditions.
A local violation of this property (e.g., in the immediate vicinity of the critical points) is allowed in our method.

A discrete version of Equation~\ref{equ:intlic} 
has already been studied in the literature \cite{Weiskopf:2009:ITLI} 
in the context of algorithm performance optimization. However, only the discrete computation of the line integral itself was taken into account.
We take the discretization a step further by explicitly specifying  discrete input and output images.

We use binary noise as input images with the same dimension as the output images to generate similar results to the output of LIC.

First, as a result of integral discretization,
we consider a finite number of particle positions,
sampled at different distances along the same streamline.
Formally, a particle $\tilde{p}$ with initial position $\mxpnew$  moving along a streamline $\mslpnew(s)$
is sampled at the distances $s_i$, 
providing a set of positions $\mxpinew = \mslpnew(s_i)$.

Next, we assume that the input image is sampled from an in-memory texture,
with some typical interpolation method (bilinear, spline) with non-negative weights,
as opposed to arbitrary procedural input texture generation
(e.g., a nonsmooth analytically defined function).
In other words,  
any input function that can be sampled on a grid and then reconstructed with interpolation
satisfies this constraint.

Suppose that the input image is given on the rectangular grid of cells $c_j$.
An input value $U(\mxpinew)$ is then interpolated as
\begin{align}
U(\mxpinew) = \sum_j u_j \mwnew \label{equ:input_linear}
\end{align}
where $u_j = U(c_j)$ is the value of the input signal at cell $c_j$, and $w_j \ge 0$ is the interpolation weight used at $c_j$. 

Now, using the interpolation formula in Equation~\ref{equ:input_linear} and the definition of $L$,
the  discrete version of the integral in Equation~\ref{equ:intlic}
can be written as 
\begin{equation}\label{equ:discrete_lic}
e_{\tilde{p}}  = \sum_{i=-L}^{L} k(t_i) \sum_j u_j \mwnew
\end{equation}
Changing the order of sums and setting 
\begin{equation*}
P_{\tilde{p}j} = \sum_{i=-L}^{L} k(t_i) \mwnew 
\end{equation*}
results in the basic matrix-vector product 
representation in 
\begin{equation*}
e_{\tilde{p}} = \sum_j P_{\tilde{p}j} u_j
\end{equation*}
Further, we call the matrix $P \in \R^{N \times N}$ a \textbf{probability matrix}, where $N$ is the number of cells, as defined in the previous section. As $k(t_i)$
and $\mwnew$ are non-negative, every element of $P$ is non-negative as well.

\subsection{Some Basic Features of the Probability Matrix}

One intriguing property of the discrete matrix formulation is that we can easily confirm the properties of other DFV methods for our operator.

\subsubsection{Filter Sequences and Iteration}
One of the techniques for the enhancement of LIC output 
is the iteration of the LIC kernel, combined with a high-pass filter
suggested by Okada and Lane \cite{Okada:1997:ELIC}.

In the matrix framework, 
the cumulative effect of a consecutive application
of the probability matrix  $P$ and another filter $B$ can be represented by
the multiplication of the input by one matrix $BP$. 
This view allows a transparent combination of our operator with basic image filters
representable in matrix form (e.g., Box, Gaussian, Laplacian).

In particular, the sequence of filters suggested by Okada and Lane
can be represented by the matrix $P^2H$,
where $H$ performs a convolution with some high-pass kernel.
Here, we do not take into account 
their final nonlinear contrast-enhancement step (histogram equalization),
which can be seen as postprocessing.

\subsubsection{Maximum and Average Value Preservation}
Additionally, we can make two basic conclusions about relating
the properties of the input and output images.
The matrix multiplication preserves the maximum norm 
(as an upper bound)
if the sum of the row elements is equal to one, i.e., $\sum_j P_{\tilde{p}j} = 1$.
Indeed, consider
\begin{align*} 
e_{\tilde{p}} = \sum_j P_{\tilde{p}j} u_j \le \sum_j P_{\tilde{p}j} \max_l u_l  = \max_l u_l
\end{align*}
The above inequality becomes an equation when
$k$ and $w$ are normalized such that,
\begin{equation*}
\sum_{i=-L}^{L} k(t_i) = 1 \quad \text{and} \quad \sum_j|\mwnew| = 1
\end{equation*}
Furthermore, the average value of the input $\mn$
is preserved under the matrix multiplication
if the sum of the matrix column elements is equal to one. 
That is, given $\sum_{\tilde{p}} P_{\tilde{p}j} = 1$ the following holds:
\begin{equation*} 
\sum_{\tilde{p}} e_{\tilde{p}} = \sum_{\tilde{p}} \sum_j P_{\tilde{p}j} u_j = \sum_j \sum_{\tilde{p}} P_{\tilde{p}j} u_j = \sum_j u_j
\end{equation*}

For a noise input, this property means, intuitively, that all pixels possess the same amount of gray value
they can redistribute among their neighbors.
In particular, the value of any pixel in the vicinity of a critical point, 
hit by multiple streamlines,
makes only a minimal contribution to the value of other pixels on each of these streamlines.
The lack of average value preservation 
can cause a visual effect of gray value smearing around critical points.

In the next section, we demonstrate how, by extending the probabilistic interpretation,
a novel flow visualization approach can be formulated. Also, we will discuss the importance
of the column-wise versus row-wise normalization
in the context of a probabilistic interpretation of the matrices.

\subsection{Idea of Particle Mixing Probability} \label{sec:dpvis}

In the probabilistic formulation in Section~\ref{probModel}, we seed a particle within each image cell
and compute the intensity value for this cell based on the trajectory of this particle.
It is then natural to switch the focus from particles directly to cells.
Such a shift would correspond to the transition 
from the Lagrangian to Eulerian approach,
common in the study of fluid dynamics.
This viewpoint change allows us to formulate the requirements for
the resulting visualization image explicitly
since the cells are directly linked to image pixels.

We are, therefore, interested in the probabilities $P(S=\tilde{p}_i|C=c_j)$.
That is,  given the cell $c_j$ is observed, 
we compute the conditional probability of 
each of the particles $\tilde{p}_i$ (each originating from its cell $c_i$)
arriving at this cell.
This cell-centric view is illustrated in \figurename~\ref{fig:particles}b. 
Applying Bayes' theorem, we get
\begin{align*}
P(S=\tilde{p}_i|C=c_j) = \frac{P(C=c_j|S=\tilde{p}_i)P(S=\tilde{p}_i)}{P(C=c_j)} 
\end{align*}
where
$P(S=\tilde{p}_i) = \sum_j P(S=\tilde{p}_i|C=c_j) $ is the marginal probability 
of a specific particle visiting any of the cells.
We assume here that for any $\tilde{p}_i$, $P(S=\tilde{p}_i) = 1$,
which is another way to say that the chosen kernel is normalized.

The marginal probability $P(C=c_j)$  that any of the observed particles visits a particular cell $c_j$ is $P(C=c_j) = \sum_i P(C=c_j|S=\tilde{p}_i)$.
Note that, technically, the transition from $P(C=c_j|S=\tilde{p}_i)$ to $P(S=\tilde{p}_i|C=c_j)$
is implemented by the normalization of columns of the original matrix.

The computed conditional probability  allows us to answer the following question:
\textbf{What is the probability that two particles $\tilde{p}_i$ and $\tilde{p}_j$ emitted from cells $c_i$ and $c_j$ 
will meet in some cell?}
By "meeting," we here refer to visiting the same cell, 
not necessarily at the same moment but within a specific interval (defined by the kernel length). 
We use $\alpha_{ij}$  to denote this probability,
which is computed  by summing up the probability of these particles
visiting the same cell $c_k$ over all cells.
\begin{equation*} 
\alpha_{ij}  = \sum_k P(S=\tilde{p}_i|C=c_k) P(S=\tilde{p}_j|C=c_k) 
\end{equation*}
Formally, this operation can be described as the computation 
of a matrix $H = P P^T$,  $H_{ij} = \alpha_{ij}$
of the conditional probability matrix $P$. 
The main diagonal of $H$, corresponding to the probability
that two particles starting at the same cell meet,
is set to 1. Since every particle originates from its own cell, $\alpha_{ij}$ describes the correlation between their cells. The probability 
that a cell $c_j$ contributes to the value of a given cell $c_i$ is therefore dependent on the used kernel (e.g., Box, Gaussian), interpolation scheme (e.g., nearest-neighbor, bilinear) and the kernel length, as well as the used step size.

This probability of two particles visiting the same cell
provides essential information about the flow domain connectivity.
It is also important to remember that the computed probabilities 
are restricted to the particle movement for a certain predefined time period.
Further, we refer to this probability as a short-term mixture probability.
Clearly, for trajectories that are nowhere closer than one cell size apart,
this probability is zero, and it is higher for largely overlapping trajectories.

\subsection{Flow Visualization Using Mixture Probability}
The short-term mixture probability
is a relation defined for every pair of cells,
that is for $N^2$ values with $N$ being the number of cells.
The direct visualization of this additional amount of data 
by itself presents a significant challenge.
However, one particular advantage of this representation is
that it allows us to compute a sequence of uncorrelated domain feature maps
that reveal the domain connectivity on different scales. 

\begin{figure*}[t!]%
\centering%
\includegraphics[width=\textwidth]{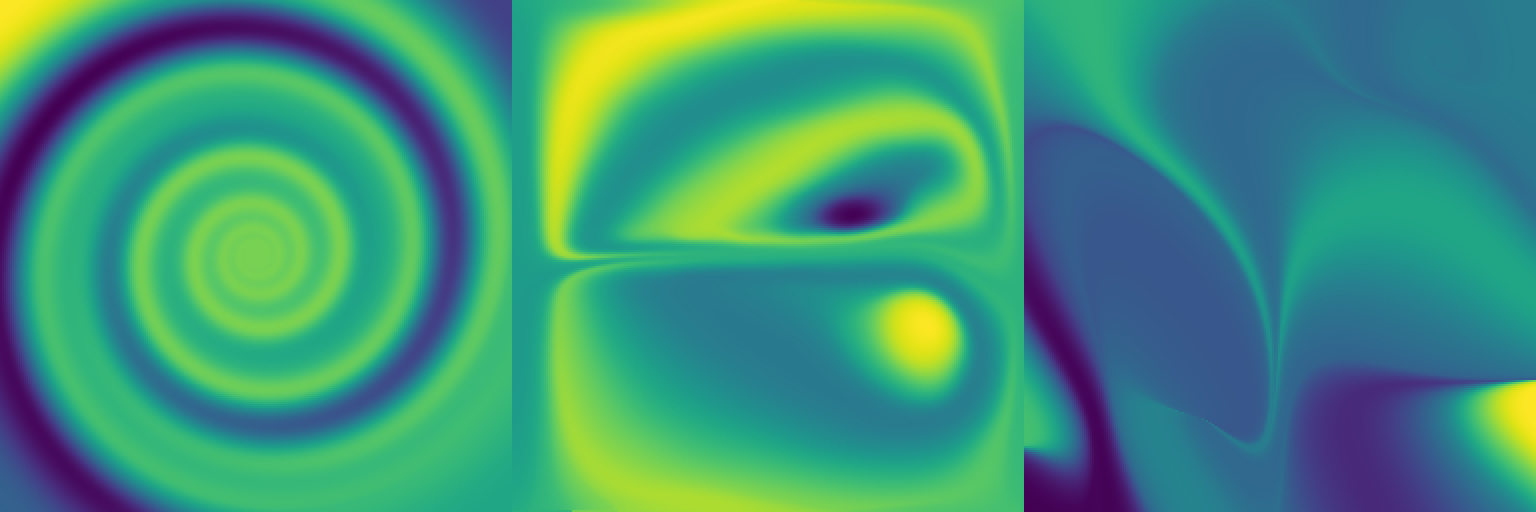}%
\caption{
A visualization of the first five embeddings for each flow combined with one transfer function.}%
\label{fig:trasnsfer_2d}%
\end{figure*}%

\begin{figure}[t!]
\centering\includegraphics[width=\columnwidth]{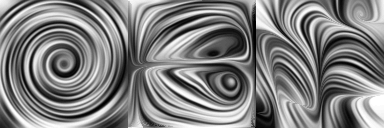}
\caption{
The underlying flow structure of the flows in \figurename~\ref{fig:trasnsfer_2d} visualized by the matrix-vector product of $P$ and a noise input vector with an additional histogram equalization step.}
\label{fig:licoutputs}
\end{figure}

We define a vector  $\mathbf f \in \R^N$ or a feature map $f_i$ for each cell $c_i$ 
as a minimizer of an expected
quadratic error function $e_{ij} = (f_i - f_j)^2$,
with constant nonzero total energy.
\begin{align} \begin{split}
E[\mathbf{f}] = \sum_{ij} \alpha_{ij} (f_i-f_j)^2\label{equ:expected_error},\\
\sum_i f_i^2 = 1 \end{split}
\end{align}

The first condition above, smoothness, ensures 
that the difference between feature map values for any two cells 
is penalized in direct proportion to their short-term mixture probability,
whereas the second condition, energy, restricts the problem to nontrivial solutions.
For the energy to be minimized, one of two things must be true: Either $\alpha_{ij}$ is small, and therefore, there is no relationship between cell $c_i$ and $c_j$, or if $\alpha_{ij}$ is large, then $(f_i - f_j)^2$ must be small. Thus, $f_i$ and $f_j$ are pushed together wherever there is a relationship between cell $c_i$ and $c_j$.

One straightforward value is achieved with a constant feature map $\mathbf{f} = \mathbf{v}^0$,
such that $v^0_i = \frac{1}{\sqrt{N}}$,
which is not useful for flow behavior analysis.
Further, we show how the complexity of the solution can be increased gradually.
The approach taken is well studied in machine learning and is
known as the spectral embedding of a graph 
(induced by the particle mixture relationship between cells). 

For two particle $\tilde{p}_i$ and $\tilde{p}_j$, a higher similarity value is assigned, if multiple points on their trajectories have only a small spatial distance between them. On the other hand, if their trajectories are far apart, $\tilde{p}_i$ and $\tilde{p}_j$ are not similar at all. This relation between particles is represented by $H$, as defined in the previous section. 

Equation~\ref{equ:expected_error} can be 
rewritten using matrix notation as a minimization of quadratic form $ E[\mathbf{f}] = 2\mathbf f^T L \mathbf f$ 
for $|| \mathbf f|| = 1$ \cite{Luxburg:2007:TSC},
where $L$ is a Laplacian matrix $L = D - H$ with 
$D$ as the degree matrix, 
holding the row sums of $H$ on the main diagonal
$D_{ii} = \sum_j H_{ij} $ and $H$ as defined in the previous section. 
By definition of $\alpha_{ij}$, L is symmetric and positive semidefinite, and therefore, we can apply the theory of spectral embeddings.
Firstly, let $\lambda_0,...,\lambda_{N-1}$ denote the eigenvalues of $L$ in ascending order, and $\mathbf{v_0},...,\mathbf{v_{N-1}}$ be the corresponding eigenvectors. It is well known from spectral embeddings, that the solution to our minimization problem is given by the first eigenvector $\mathbf{v_0}$ of $L$, minimizing $E[\mathbf{f}]$ with value $2\lambda_0$.

 \begin{figure*}[t!]%
\centering
\includegraphics[width=\textwidth]{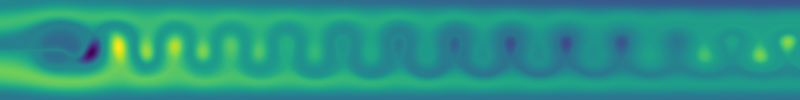}
\caption{
A visualization of the combination of the first five embeddings for the von K\'arm\'an vortex street in the flow behind a cylinder. }
\label{fig:ftleCompare}
\end{figure*}

As mentioned, the first eigenvector is of little interest.
However, introducing the additional constraint, for the new solution to be orthogonal to the constant one, we can restrict the space of solutions to non-constant vectors.
With this constraint, the new solution is given by $\mathbf{v_1}$ with $2\lambda_1$ as the minimal value.

Repeating this step, we progressively refine the vector space of admitted solutions, 
constructing increasingly more detailed feature maps,
which we call \textbf{flow spectral embeddings} by analogy with graph theory.
By requiring the new solution to be orthogonal to the previously computed solution, 
we ensure that the new image and any linear combination of previously computed images are not correlated.

At the same time, the smoothness objective 
controls the discrepancy in the solution cell values
according to the cell mixture probability
so that cells with a higher probability of mixing
maintain a small difference in value for a larger number of steps.

Analytically the kernel of the Laplacian can have a dimension larger than one, corresponding to disconnected areas. Therefore, the eigenvalue 0 would have multiple eigenvectors. Numerically this case is present in unstructured grids with omitted parts in the domain. As we currently cannot handle unstructured grids, a kernel of dimension larger than one presents a limitation of our algorithm. 

The described algorithm computes a sequence of feature maps,
by visualizing the short-term cell mixture probability in the domain
and increasingly adding small features.
We can summarize the steps of the algorithm as follows:
\begin{enumerate}
\item Choose the integration distance, that is the kernel length $L$ and the step size $s$, and the trajectory certainty distribution, the kernel function.
\item Trace a particle $\tilde{p}_i$ for each cell $c_i$.
\item Sample the particle positions and store the probability $P(c_j|\tilde{p}_i)$ of visiting cell $c_j$, 
given the particle $\tilde{p}_i$ is observed.
\item Store $P(c_j|\tilde{p}_i)$ in a matrix $P$ and normalize $P$ by rows.
\item Compute the probability $P(\tilde{p}_i|c_j)$ of particle $\tilde{p}_i$ visiting the cell $c_j$,
normalizing the matrix $P$ by columns.
\item Compute the short-term mixing probability matrix $H = P P^T $, i.e., 
the probability that two particles, originating from cells $c_i$ and $c_j$, meet at some cell.
\item Compute the Laplacian matrix $L =  D - H$.
\item Compute first several eigenvectors of the Laplacian matrix (corresponding to smallest eigenvalues).
\end{enumerate}

For visualization purposes, we explore a few of the first such maps (embeddings),
that is, eigenvectors corresponding to 
the first several smallest eigenvalues of the Laplacian matrix.
The obtained sequence of images exhibits a multiscale structure,
with the spatial frequency increasing with the number of eigenvectors
computed, as demonstrated by the top row in \figurename~\ref{fig:teaser}. In the background of the top row of \figurename~\ref{fig:teaser}, the result of the matrix-vector product between our probability matrix and an input noise vector is added.
It is important to note that the maps are ordered by the expected error and are orthogonal to each other.
In other words, each subsequent map contains a smaller amount of detail
compared to the previously constructed set.
This detail is not present in the previous maps and has a higher spatial frequency (lower smoothness). 
Consequently, the eigenvectors with large indices (specific number depending on the flow) 
appear to be uniform noise, which is a result of the spatial frequency reaching the Nyquist limit of the chosen image resolution 
and the increasing numerical imprecision associated with eigenvalue computation.

The resulting images
visualize the connectivity of the domain based on flow transport,
since the mixture probability for any two small domain regions, 
which is defined as the probability of the meeting of particles seeded in these regions,
corresponds to the difference in colors assigned to these regions in the visualization.
In our view, such a presentation simplifies the interpretation 
of the flow mixture patterns even for a non-expert user.

Although spectral embeddings by themselves 
already capture the flow structure
and, to some extent, visualize the flow transport,
they can be combined into a single image 
using a transfer function for
enhancement of detail on different scales.
The examples of such visualization for the 2D case are shown in \figurename~\ref{fig:trasnsfer_2d} and \figurename~\ref{fig:ftleCompare}. The underlying flow structure can be seen in \figurename~\ref{fig:licoutputs}.
In this case, we construct the transfer function as
\begin{equation*}
TF=c\left(\sum_k \frac{a_k}{\sum_l a_l} v_k\right)
\end{equation*}
where $v_k$ are the embeddings, $a_k$ the respective amplitudes, and $c$ the chosen colormap. In the next section, we see how we can estimate the amplitudes for our eigenvectors. 

The results in \figurename~\ref{fig:trasnsfer_2d} and \figurename~\ref{fig:ftleCompare} are not supported by other visualization techniques to present the results itself. However, multiplying $P$ with a noise input vector results in outputs similar to LIC nearly without additional overhead. Therefore, we suggest using the presented results with this additional step to get a better grasp of the underlying flow structure.
The practical computation of the embeddings is discussed in Section~\ref{sec:implementation}.
In summary, the numerical scheme requires only a series of sparse matrix-vector multiplications,
which are provided by many general-purpose 
sparse  linear algebra software packages,
including those that are GPU-accelerated. Examples are SPARSPAK \cite{SPARSPAK}, the Yale Sparse Matrix Package \cite{YalePackage}, and for GPU-accelerated packages cuSPARSE.

The relation between our method and spectral embeddings in general to frequencies can be seen from the following consideration.
 The idea of spectral embedding is to minimize the objective function $E[\mathbf{f}]$ to get coherent structures in the data \cite{Luxburg:2007:TSC}. The solution to this minimization is the first eigenvector. However, as the first eigenvector is constant, the data are not separated in multiple coherent structures. Therefore, the orthogonality constraint is used to separate the flow into more and more coherent structures. The solutions to this new objective function are the eigenvectors of the Laplacian matrix in increasing order. If we go from eigenvectors corresponding to smaller eigenvalues to eigenvectors corresponding to higher eigenvalues, we cut our graph into increasingly smaller coherent structures with an overall higher similarity. This phenomenon follows from the idea that the Laplacian matrix is a discretization of the Laplace operator on graphs. The eigenvectors of the Laplacian matrix correspond to a discretization of the eigenfunctions of the continuous Laplace operator. These eigenfunctions are the terms of a Fourier expansion. Therefore, eigenvectors associated with small eigenvalues are "smooth, slowly oscillating," and eigenvectors associated with bigger eigenvalues "oscillate much more rapidly" \cite{LaplacianEigenvalueFrequency}.

\textbf{Now, what can we observe in flow spectral embeddings?} We use the eigenvectors of the Laplacian to visualize coherent structures in the flow, which can be seen in the top row of \figurename~\ref{fig:teaser}. The first image is separated into two coherent structures: the top and bottom halves. Points in the flow that are connected by a straight line with only marginal changes in color have a higher coherence than points connected by a line with significant changes in color. The second image is separated into more coherent structures (the surrounding yellowish and the two enclosed "oval" structures), which exhibit an overall higher coherence between points in them. In the third and fourth images, the flow gets separated even further with an increasing similarity of streamlines in those structures. 

However, the color does not convey any additional information about the flow. That means disconnected regions with the same color have nothing in common. They only have the same color by coincidence, since there is only a limited amount of different colors in the colormap. For example, in the second image in the top row of \figurename~\ref{fig:teaser}, the two vortices have the same color. The flow in each vortex is similar, as the trajectories overlap. Despite that, the vortices itself are not similar, as there is no overlap between their trajectories.

\subsection{Eigenvector Selection}

We would like to have a measure to differentiate between eigenvectors that capture more important structures and eigenvectors that capture less important structures to a specific flow. A global statement for this cannot be given without calculating all eigenvectors. It is possible, however, to locally determine the importance of the calculated eigenvectors. As previously stated, the eigenvalues of the Laplacian matrix correspond to frequencies, and the eigenvectors are a discrete version of the terms of a Fourier expansion. Therefore, we determine the amplitudes for the eigenvectors and decide which are essential for the flow and which can be neglected. The amplitudes of the eigenfunctions of the Laplace operator can be characterized by the $L^p$ norms ($p \ge 1$) \cite{LaplacianEigenvectorAmplitude}
As the eigenvectors are discrete evaluations of the eigenfunctions, it is natural to characterize the amplitudes of the eigenvectors with the $p$ norms ($p \ge 1$)
\begin{equation*}
||\mathbf{u}||_p = \left(\sum_i |u_i|^p\right)^{\frac{1}{p}}, \qquad ||\mathbf{u}||_{\infty} = \max_i |u_i|
\end{equation*}

The eigenvector selection can be summarized as follows: Calculate the amplitudes for $k$ given eigenvectors, sort them in descending order of the amplitudes, and use the first $1 \le m \le k$ eigenvectors to composite the final image.

\subsection{Method Extensions}
  
 \begin{table*}[t!]\centering
\caption{Run-time measurements for computation of the embeddings for the 2D flows given in \figurename~\ref{fig:trasnsfer_2d} and \figurename~\ref{fig:ftleCompare}, as well as some of the presented 3D flows. }
\label{tab:perf}
\begin{tabular}{lccrr} 
\bf{flow}          & 
\bf{output resolution} & 
\bf{half kernel length} &
\bf{timings(minutes) } \\ \midrule

\figurename~\ref{fig:trasnsfer_2d}.a    & $256\times 256$       &  100  & 85 \\
\figurename~\ref{fig:trasnsfer_2d}.b    & $256\times 256$         &  50 & 91 \\
\figurename~\ref{fig:trasnsfer_2d}.c    & $256\times 256$       &  100  & 83 \\
\figurename~\ref{fig:ftleCompare}   & $800\times 100$     &  100  & 76 \\
borromean        & $128 \times 128 \times 128$  & 80  & 149 \\                 
abc              & $128 \times 128 \times 128$  & 80  & 118 \\    
benard           & $128 \times 128 \times 128$  & 80  & 198 \\
spherical drop   & $128 \times 128 \times 128$  & 80  & 106 \\ 
stuart vortex    & $128 \times 128 \times 128$  & 80  & 113 \\
\end{tabular}
\end{table*}

The probabilistic model we described consists 
of several pluggable components (the chosen kernel, interpolation, and integration scheme),
which can be adjusted to the requirements
for a particular visualization task at hand.
These settings are not all specific to our approach,
but they are widely shared with the LIC method.  
We interpret the effect of possible modifications in the new context,
not aiming to demonstrate all their combinations.
Staying within the bounds of the expected error minimization framework,
we briefly discuss different approaches for the definition 
of the input probabilities and different domain types.

The trajectory certainty distribution corresponds to the kernel
with the restriction that it represents the conditional probability and 
hence is positive and normalized.
We make no assumptions about the shape of the kernel; in particular, it does not have to be symmetric.  
For example, a one-sided kernel (defined on a positive or negative half of the real line)
would correspond to either injecting particles in the cell and following their position
or registering the arriving particles' trajectories in the cell.
The mixing probability then can be interpreted 
as the probability of mixing in the future (reaching a common sink)
or in the past (originating from a common source).
Since the short-time mixing probability matrix properties 
are independent of the kernel shape,  
the other computational steps remain unchanged.

Different objectives can be achieved with two alternative types of 
trajectory parameterization:
by time or by distance traveled by a particle. 
In the first case, the mixing probability corresponds to mixing in a particular period.
In the second case, mixing in a neighborhood of a particular size in space 
is considered,  effectively ignoring the particle velocity magnitude.

Finally, the presented method does not make any assumptions 
about the dimensionality of the flow domain
and hence is applicable in 3D without modification.
The straightforward extension of the method to 3D makes it possible to use standard
volume rendering utilities to visualize the flow embeddings.
We demonstrate several 3D examples in Section~\ref{sec:emb3d}.

\section{Implementation Details} \label{sec:implementation}  

In this section, we highlight some technical aspects of our
current implementation and mention possible performance issues.
The core of the method is the computation of 
eigenvectors of the Laplacian matrix $L$.
The computation of the conditional probability matrix $P$ entries   
is not substantially different from the standard LIC algorithm.
Moreover, the search for eigenvectors relies on well-established algebraic routines,
which allows for straightforward implementation 
using existing software packages for scientific computation, such as ARPACK \cite{ARPACK}, PETSc \cite{PETSc}, and Eigen \cite{Eigen}.

The subsequent  steps are simplified by the  explicit storage 
of uncertainties $P(c_j|\tilde{p}_i)$ in a matrix.
However, it is not necessary to keep this matrix in memory
since the probability $P(\tilde{p}_i|c_j)$ can be computed directly
by summing up the contribution of all incoming particles for each cell.

We implemented the whole process in C++, solely on the CPU. The integral line segments are computed using the classical Runge-Kutta integration scheme with bilinear interpolation and a Gaussian kernel. 

Encouraging interested readers to reproduce the presented results, we provide the following pseudo-code of the described algorithm for the 2D case. 

\begin{lstlisting}[language=C++]
void calcSpectralEmbeddings() {
    // creates the probability matrix from 
    // the given data
    P = calcProbabilityMatrix();
    
    P.normalizeByRowsum();
    
    if (calculate LIC output) {
        // generates a random vector with entries 
        // in {0, 1}
        binaryNoise = generateNoise();
        licOutput = P * binaryNoise;
        display(licOutput);
    }
    
    P.normalizeByColsum();
    
    H = P * transpose(P);
    // creates a diagonal matrix with the sum of 
    // every row of H as the entries
    D = diagonalMatrix(rowsum(H));
    L = D - H;
    
    // calculate the smallest numberEigs 
    // eigenvectors with Lanczos
    eigenvectors = Lanczos(L, numberEigs);
    
    display(eigenvectors);
}

\end{lstlisting}

The 2D datasets can be found in \cite{KruegerDataset}. The von K\'arm\'an vortex street in the flow behind a cylinder has been simulated using the Free Software \emph{Gerris Flow Solver} \cite{gerrisflowsolver}. A related experiment for three interlocked magnetic Borromean flux rings \cite{Candelaresi2011a} is shown in \figurename~\ref{fig:teaser}.
The Stuart vortex can be described by a closed formula:
\begin{equation*}
    \left(\sinh(2y),\text{ }\frac{1}{4}\sin(2(t-x),\text{ } z(\cosh(2y)-\frac{1}{4}\cos(2(t-x))\right)
\end{equation*}

The Benard data set was obtained using the software NaSt3DGP, developed by Institute for Numerical Simulation, all rights Institute for Numerical Simulation, University of Bonn \cite{numSim1, numSim2}. For the ABC flow, see the typically used definition in \cite{ABC}.

\subsection{Sparse Matrix Computations}
  
A fundamental property of our matrices that makes the eigenvector computation feasible in practice is their sparsity.
Let us assume the average number of pixels or voxels $D$ on each particle trace 
is much smaller than the image/volume domain size $N$ (the total number of pixels/voxels),
and the interpolation requires $K$ input pixels to compute one output pixel.
 
By definition of $H$, matrix entries are nonzero, only if the trajectories of both particles share mutual cells. Consequently, only particles that are not further apart than $D$ cells from each other can have nonzero matrix entries. So, for one specific cell, the maximum number of cells with overlapping trajectories is given by
\begin{equation*}
    2\sum_{i=1}^{D-1}(2i-1)+2D-1 = 2D^2-2D+1
\end{equation*}
Therefore, the number of nonzero values in each row of $H$ grows in relation to $O(KD^2)$. With $H$ containing $N$ rows, the total number of nonzero values is $O(NKD^2)$.

As $L$ only differs on its main diagonal from $H$, setting every entry to zero, the sparse complexity differs in the subtrahend $N$. Hence, the sparse complexity of $L$ is $O(NKD^2)$ as well.

This estimation is the upper bound for the worst case, where every cell is connected to every other cell in its neighborhood. In practice, the number of nonzero entries should be many factors lower.

Since the matrix size grows as $O(N^2)$ with the size of the image or volume $N$, naively handling large domains on a computer becomes quickly challenging. 
Fortunately, several numerical methods are available that exploit the sparsity structure, 
featuring memory requirements and run times that are linear in the number of nonzero entries, such as the locally optimal block PCG method (LOBPCG) \cite{Knyazev:2001:TOP:587151.587187}.

\subsection{Performance}
At the moment, our implementation is purely prototypical 
and requires significant preprocessing times.    
For 2D images, the run time of our method is  
dominated by the probability matrix computation, 
whereas the visualization image itself is formed in order of seconds. However, as our probabilistic matrix computation is dependent on the calculation of streamlines similar to LIC, we can make use of optimized LIC algorithms.
The eigenvector computation, however, 
is the current bottleneck in 3D 
and can last from several minutes to several hours.
It is vital to notice
that this step involves only sparse matrix-vector multiplications 
and can be parallelized effectively on multi-/many-core hardware
using, for example, the several available libraries for
sparse matrix computations on GPU, such as cuSPARSE.  

Generally, the computation time can be affected by many factors,
including 
\begin{itemize} 
  \item output image/volume resolution,
  \item kernel length,
  \item number of computed eigenvectors,        
  \item interpolation scheme,
  \item sparse matrix storage scheme. 
\end{itemize}

From the theoretical perspective, the memory requirements
of our technique and asymptotic complexity 
of the involved algorithms are both linear 
in the number of non-zero entries of the probability matrix.

Actual timings for the 2D flows in \figurename~\ref{fig:trasnsfer_2d} and \figurename~\ref{fig:ftleCompare}, as well as some of the presented 3D flows, are provided in Table~\ref{tab:perf}.

\subsection{Embeddings in 3D} \label{sec:emb3d}

    \begin{figure}[t]   \centering
  \includegraphics[width=\columnwidth]{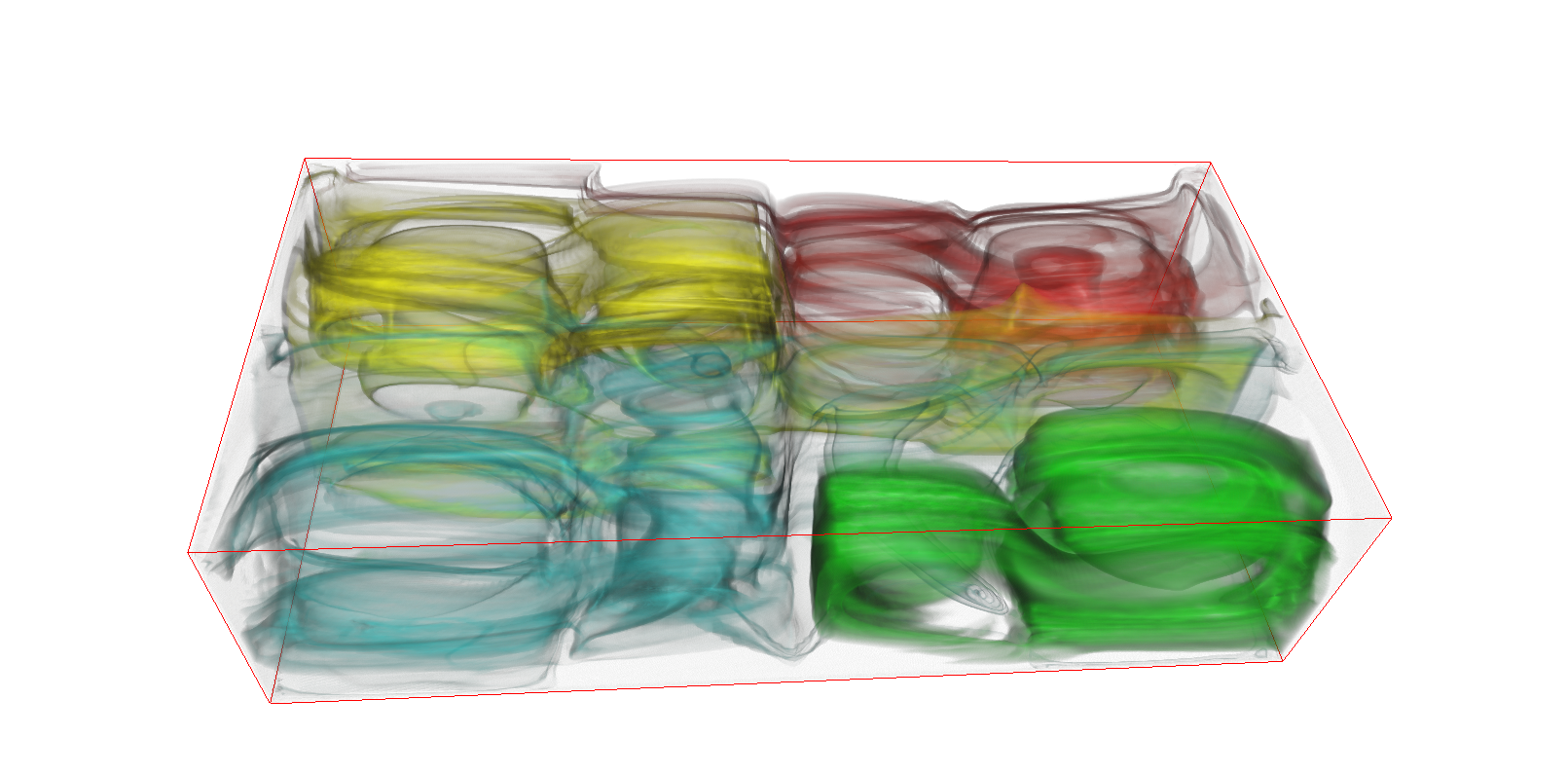} \caption{
    Direct volume rendering of first two  spectral embeddings
    combined with a color transfer function $TF$ for Benard flow.    
  }
  \label{fig:benard}%
  \end{figure}

  The flow spectral embeddings are themselves scalar fields, 
  intuitively featuring the separation of the domain
  into regions with low intermixing between them.
  Since these scalar fields are computed for each cell of the entire (discrete) domain, 
  they are volumes rather than surfaces and can be visualized using known 
  volume rendering methods and techniques.
  Unlike traditional dense flow visualization techniques,
  no dense noise is present, and 
  certain flow regions can be highlighted 
  with a modification of basic transfer function properties
  such as transparency and color.
  In other words, what makes this particular visualization method
  attractive in 3D is that a local region of interest in the flow
  corresponds to a certain bounded continuous 
  range in the embedding scalar field values,
  which makes it easy to select and filter.    
  Therefore, we cluster the scalar field values into bounded regions with thresholds where the values jump. We then assign a specific color to every region. This clustering results in a separation of the flow into regions where particles interact with each other strongly. The interaction between these regions is low. By setting the transparency to 1, filtering the regions that cover large parts of the domain and are of no interest is easily done. The result is a visualization of the flow into regions with low intermixing between them, as shown in the presented results of 3D flows, such as the bottom row of \figurename~\ref{fig:teaser}.

 \begin{figure}[t]   \centering
  \includegraphics[width=\columnwidth]{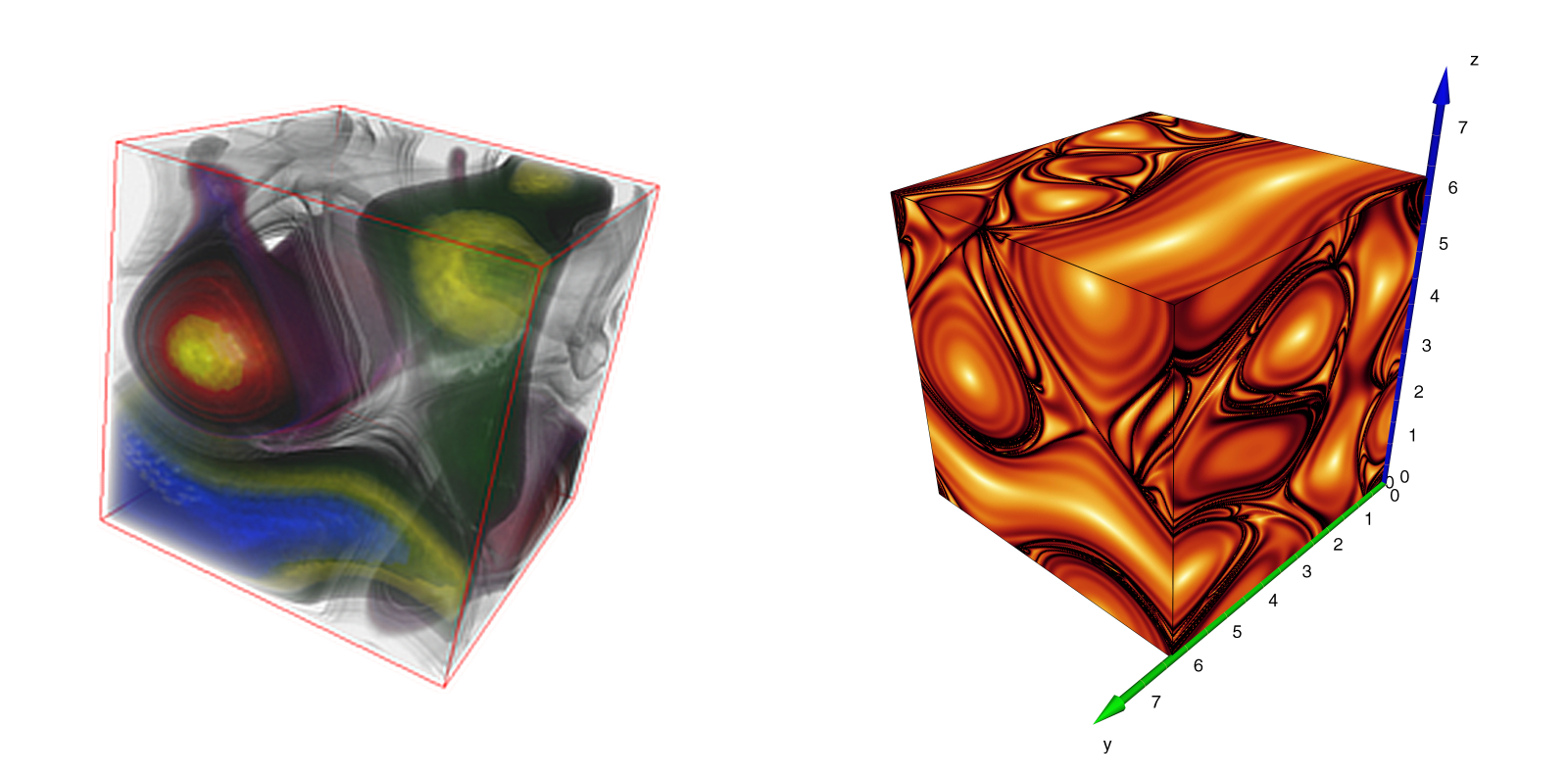} \caption{
    Direct volume rendering of the first spectral embedding for ABC flow (left),    
    compared to the Lyapunov exponential, 
    reproducing previous visualization of Haller \cite{Haller:2001:DMCS} for the same flow.    
  }
\label{fig:abc}%
  \end{figure}
  
  Using the transfer function $TF$ with the above-described coloring, \figurename~\ref{fig:benard} reveals four compartments of the Benard flow.
  The symmetry of the dataset is well captured. 
  
  \figurename~\ref{fig:abc} demonstrates some of the structures in the embedding volume 
  for the ABC flow. Remarkably, the contours of the structures are similar to  
  the ridges of the Lyapunov exponential map,
  shown in the reproduced visualization of Haller \cite{Haller:2001:DMCS}.

\section{Future Work}

In this work, we have presented a novel formal approach
in dense flow visualization
based on matrix and probability theory. 
The consequences of this probabilistic model
lead to a novel dense flow visualization technique:
flow spectral embeddings.
We consider the demonstrated results of this method
as valuable for visualization practitioners 
and encouraging for further development of the underlying ideas.

At present, the discussed techniques are verified 
with a proof-of-concept prototypical implementation,
which requires an optimization effort to reach the production software standards.
The refinement of technical aspects of the method is another research direction  we plan to take in the future.
In particular, the focus on the multiscale aspect of the eigenvector representation of the flow domain is of interest.
The concept of particle trajectories is generic and is not limited to the physical particle movement.
In the case of unsteady flow, the probability of virtual particles mixing in the domain cells can be computed for
streamlines as well as streaklines or pathlines.
Therefore, it could be of interest to investigate the robustness of the embeddings under flow modification and their insightfulness.
For the 3D case, streamline clustering to reveal more compelling regions could be of interest.

On the technical side, we plan to adopt a straightforward parallel implementation 
for assembling the probability matrix, as well as the eigenvector computation on GPU,
which is expected to decrease the run times dramatically.
The systematic theoretical treatment of the eigenvector computation might 
further speed up the computation employing
effective preconditioners in combination with the LOBPCG algorithm.

Furthermore, we plan to analyze the scalability of our method in the future. Increasing the size of the output, and respectively, the size of the matrices, the density of the matrices gets smaller. Coupled with the computational complexity that is linear in the number of nonzero entries of the LOBPCG method, applying our algorithm to larger data seems promising.



\bibliographystyle{IEEEtran}
\bibliography{IEEEabrv,paper}

\begin{thebibliography}{10}
\providecommand{\url}[1]{#1}
\csname url@samestyle\endcsname
\providecommand{\newblock}{\relax}
\providecommand{\bibinfo}[2]{#2}
\providecommand{\BIBentrySTDinterwordspacing}{\spaceskip=0pt\relax}
\providecommand{\BIBentryALTinterwordstretchfactor}{4}
\providecommand{\BIBentryALTinterwordspacing}{\spaceskip=\fontdimen2\font plus
\BIBentryALTinterwordstretchfactor\fontdimen3\font minus
  \fontdimen4\font\relax}
\providecommand{\BIBforeignlanguage}[2]{{%
\expandafter\ifx\csname l@#1\endcsname\relax
\typeout{** WARNING: IEEEtran.bst: No hyphenation pattern has been}%
\typeout{** loaded for the language `#1'. Using the pattern for}%
\typeout{** the default language instead.}%
\else
\language=\csname l@#1\endcsname
\fi
#2}}
\providecommand{\BIBdecl}{\relax}
\BIBdecl

\bibitem{Cabral:1993:IVFU}
\BIBentryALTinterwordspacing
B.~Cabral and L.~C. Leedom, ``Imaging vector fields using line integral
  convolution,'' in \emph{Proceedings of the 20th Annual Conference on Computer
  Graphics and Interactive Techniques}, ser. SIGGRAPH '93.\hskip 1em plus 0.5em
  minus 0.4em\relax New York, NY, USA: ACM, 1993, pp. 263--270. [Online].
  Available: \url{http://doi.acm.org/10.1145/166117.166151}
\BIBentrySTDinterwordspacing

\bibitem{stalling95}
D.~Stalling and H.-C. Hege, ``Fast and resolution independent line integral
  convolution,'' \emph{Proceedings Siggraph '95}, pp. 249--256, 1995, los
  Angeles.

\bibitem{Wijk:1991:SNTS}
\BIBentryALTinterwordspacing
J.~J. van Wijk, ``Spot noise texture synthesis for data visualization,''
  \emph{SIGGRAPH Comput. Graph.}, vol.~25, pp. 309--318, July 1991. [Online].
  Available: \url{http://doi.acm.org/10.1145/127719.122751}
\BIBentrySTDinterwordspacing

\bibitem{vanWijk:imagebasedflowvis}
\BIBentryALTinterwordspacing
------, ``Image based flow visualization,'' \emph{ACM Trans. Graph.}, vol.~21,
  no.~3, pp. 745--754, Jul. 2002. [Online]. Available:
  \url{http://doi.acm.org/10.1145/566654.566646}
\BIBentrySTDinterwordspacing

\bibitem{Laramee:2004:TSOA}
\BIBentryALTinterwordspacing
R.~S. Laramee, H.~Hauser, H.~Doleisch, B.~Vrolijk, F.~H. Post, and D.~Weiskopf,
  ``The state of the art in flow visualization: Dense and texture-based
  techniques,'' \emph{Computer Graphics Forum}, vol.~23, pp. 203--221, 2004.
  [Online]. Available:
  \url{http://www.winslam.com/rlaramee/star/post03state.pdf}
\BIBentrySTDinterwordspacing

\bibitem{Brand:2003:UTSEC}
M.~Brand and K.~Huang, ``A unifying theorem for spectral embedding and
  clustering,'' in \emph{AISTATS}, 2003.

\bibitem{Luxburg:2007:TSC}
\BIBentryALTinterwordspacing
U.~Luxburg, ``A tutorial on spectral clustering,'' \emph{Statistics and
  Computing}, vol.~17, no.~4, pp. 395--416, Dec. 2007. [Online]. Available:
  \url{http://dx.doi.org/10.1007/s11222-007-9033-z}
\BIBentrySTDinterwordspacing

\bibitem{Rezk-Salama:1999:IEOV}
\BIBentryALTinterwordspacing
C.~Rezk-Salama, P.~Hastreiter, C.~Teitzel, and T.~Ertl, ``Interactive
  exploration of volume line integral convolution based on 3d-texture
  mapping,'' in \emph{Proceedings of the conference on Visualization '99:
  celebrating ten years}, ser. VIS '99.\hskip 1em plus 0.5em minus 0.4em\relax
  Los Alamitos, CA, USA: IEEE Computer Society Press, 1999, pp. 233--240.
  [Online]. Available: \url{http://dl.acm.org/citation.cfm?id=319351.319379}
\BIBentrySTDinterwordspacing

\bibitem{Forssell:1994:VFCG}
\BIBentryALTinterwordspacing
L.~K. Forssell, ``Visualizing flow over curvilinear grid surfaces using line
  integral convolution,'' in \emph{Proceedings of the conference on
  Visualization '94}, ser. VIS '94.\hskip 1em plus 0.5em minus 0.4em\relax Los
  Alamitos, CA, USA: IEEE Computer Society Press, 1994, pp. 240--247. [Online].
  Available: \url{http://dl.acm.org/citation.cfm?id=951087.951132}
\BIBentrySTDinterwordspacing

\bibitem{UFLIC}
H.-W. Shen and D.~L. {Kao}, ``A new line integral convolution algorithm for
  visualizing time-varying flow fields,'' \emph{IEEE Transactions on
  Visualization and Computer Graphics}, vol.~4, no.~2, pp. 98--108, April 1998.

\bibitem{Stalling:1995:FARI}
\BIBentryALTinterwordspacing
D.~Stalling and H.-C. Hege, ``Fast and resolution independent line integral
  convolution,'' in \emph{Proceedings of the 22nd Annual Conference on Computer
  Graphics and Interactive Techniques}, ser. SIGGRAPH '95.\hskip 1em plus 0.5em
  minus 0.4em\relax New York, NY, USA: ACM, 1995, pp. 249--256. [Online].
  Available: \url{http://doi.acm.org/10.1145/218380.218448}
\BIBentrySTDinterwordspacing

\bibitem{Weiskopf:2006:GPUB}
\BIBentryALTinterwordspacing
D.~Weiskopf, \emph{GPU-Based Interactive Visualization Techniques (Mathematics
  and Visualization)}.\hskip 1em plus 0.5em minus 0.4em\relax Secaucus, NJ,
  USA: Springer-Verlag New York, Inc., 2006. [Online]. Available:
  \url{http://www.springer.com/mathematics/numerical+and+computational+mathematics/book/978-3-540-33262-6}
\BIBentrySTDinterwordspacing

\bibitem{Stalling:1998:FTBBA}
\BIBentryALTinterwordspacing
D.~Stalling, ``Fast texture-based algorithms for vector field visualization,''
  Ph.D. dissertation, Zuse Institute Berlin, 1998. [Online]. Available:
  \url{http://vs24.kobv.de/documents-zib/383/SC-98-40.pdf}
\BIBentrySTDinterwordspacing

\bibitem{Okada:1997:ELIC}
\BIBentryALTinterwordspacing
A.~Okada and D.~Lane, ``Enhanced line integral convolution with flow feature
  detection,'' in \emph{SPIE Vol. 3017 Visual Data Exploration and Analysis
  IV}, 1997, pp. 206--217. [Online]. Available:
  \url{http://dx.doi.org/10.1117/12.270314}
\BIBentrySTDinterwordspacing

\bibitem{Weiskopf:2009:ITLI}
\BIBentryALTinterwordspacing
D.~Weiskopf, ``Iterative twofold line integral convolution for texture-based
  vector field visualization,'' in \emph{Mathematical Foundations of Scientific
  Visualization, Computer Graphics, and Massive Data Exploration}, ser.
  Mathematics and Visualization.\hskip 1em plus 0.5em minus 0.4em\relax
  Springer Berlin Heidelberg, 2009, pp. 191--211. [Online]. Available:
  \url{http://dx.doi.org/10.1007/b106657_10}
\BIBentrySTDinterwordspacing

\bibitem{Hlawatsch:2010:HLI}
\BIBentryALTinterwordspacing
M.~Hlawatsch, F.~Sadlo, and D.~Weiskopf, ``Hierarchical line integration,''
  \emph{IEEE Transactions on Visualization and Computer Graphics}, vol.~99,
  2010. [Online]. Available:
  \url{http://doi.ieeecomputersociety.org/10.1109/TVCG.2010.227}
\BIBentrySTDinterwordspacing

\bibitem{Matvienko:2013:AMFT}
\BIBentryALTinterwordspacing
V.~Matvienko and J.~Kr{\"u}ger, ``A metric for the evaluation of dense vector
  field visualizations,'' \emph{IEEE Transactions on Visualization and Computer
  Graphics,}, vol.~19, no.~7, pp. 1122--1132, 7 2013. [Online]. Available:
  \url{http://www.ivda.uni-saarland.de/fileadmin/publications/2012/tvcg\_metric.pdf}
\BIBentrySTDinterwordspacing

\bibitem{Reich:2012:ASSI}
\BIBentryALTinterwordspacing
W.~Reich and G.~Scheuermann, ``Analysis of streamline separation at infinity
  using time-discrete markov chains,'' \emph{IEEE Transactions on Visualization
  and Computer Graphics}, vol.~18, p.~9, 2012. [Online]. Available:
  \url{http://www.informatik.uni-leipzig.de/~reich/vis12.pdf}
\BIBentrySTDinterwordspacing

\bibitem{Salzbrunn:2008:STPT}
T.~Salzbrunn, T.~Wischgoll, H.~JÃ¤nicke, and G.~Scheuermann, ``The state of
  the art in flow visualization: Partition-based techniques,'' in \emph{In
  Simulation and Visualization 2008 Proceedings}, 2008.

\bibitem{Salzbrunn:2008:TSOA}
\BIBentryALTinterwordspacing
T.~Salzbrunn, H.~J{\"a}nicke, T.~Wischgoll, and G.~Scheuermann, ``The state of
  the art in flow visualization: Partition-based techniques,'' in
  \emph{SimVis}, 2008, pp. 75--92. [Online]. Available:
  \url{http://cs.swan.ac.uk/~csheike/Data/simVis08.pdf}
\BIBentrySTDinterwordspacing

\bibitem{helman89}
J.~L. Helman and L.~Hesselink, ``Representation and display of vector field
  topology in fluid flow data sets,'' \emph{IEEE Computer}, vol.~22, no.~8, pp.
  27--36, August 1989.

\bibitem{tricoche01a}
X.~Tricoche, G.~Scheuermann, and H.~Hagen, ``Continuous topology simplification
  of planar vector fields,'' in \emph{Proc.\ Visualization}, 2001, pp. 159 --
  166.

\bibitem{weinkauf05a}
T.~Weinkauf, H.~Theisel, K.~Shi, H.-C. Hege, and H.-P. Seidel, ``Extracting
  higher order critical points and topological simplification of 3d vector
  fields,'' in \emph{Proc.\ IEEE Visualization 2005}, 2005, pp. 559--566.

\bibitem{reininghaus11a}
J.~Reininghaus and I.~Hotz, ``Combinatorial {2D} vector field topology
  extraction and simplification,'' in \emph{Topological Methods in Data
  Analysis and Visualization}, ser. Mathematics and Visualization, V.~Pascucci,
  X.~Tricoche, H.~Hagen, and J.~Tierny, Eds.\hskip 1em plus 0.5em minus
  0.4em\relax Springer, 2011, pp. 103--114.

\bibitem{Szymczak:2013:VMCG}
A.~Szymczak and L.~Sipeki, ``Visualization of morse connection graphs for
  topologically rich 2d vector fields,'' \emph{IEEE Trans. Vis. Comput.
  Graph.}, vol.~19, no.~12, pp. 2763--2772, 2013.

\bibitem{efficientMorse}
G.~{Chen}, K.~{Mischaikow}, R.~S. {Laramee}, and E.~{Zhang}, ``Efficient morse
  decompositions of vector fields,'' \emph{IEEE Transactions on Visualization
  and Computer Graphics}, vol.~14, no.~4, pp. 848--862, July 2008.

\bibitem{Salzbrunn:2006:SP}
T.~Salzbrunn and G.~Scheuermann, ``Streamline predicates,'' \emph{IEEE
  Transactions on Visualization and Computer Graphics}, vol.~12, no.~6, pp.
  1601--1612, 2006.

\bibitem{Peng}
Z.~{Peng}, E.~{Grundy}, R.~S. {Laramee}, G.~{Chen}, and N.~{Croft},
  ``Mesh-driven vector field clustering and visualization: An image-based
  approach,'' \emph{IEEE Transactions on Visualization and Computer Graphics},
  vol.~18, no.~2, pp. 283--298, Feb 2012.

\bibitem{Roessl:2012:TVCG}
\BIBentryALTinterwordspacing
C.~R{\"o}ssl and H.~Theisel, ``Streamline embedding for 3d vector field
  exploration,'' \emph{IEEE Transactions on Visualiztion and Computer
  Graphics}, vol. 18-3, pp. 407--420, 2012, fast track TVCG from IEEE
  Visualization 2010. [Online]. Available:
  \url{http://wwwisg.uni-magdeburg.de/visual}
\BIBentrySTDinterwordspacing

\bibitem{Garcke:2002:CCMVF}
H.~Garcke, T.~Preusser, M.~Rumpf, A.~Telea, U.~Weikard, and J.~J. van Wijk, ``A
  continuous clustering method for vector fields.'' in \emph{IEEE
  Visualization}, 2002, pp. 351--358.

\bibitem{heckel99}
B.~Heckel, G.~Weber, B.~Hamann, and K.~I. Joy, ``Construction of vector field
  hierarchies,'' in \emph{Proc.\ IEEE Visualization '99}, D.~Ebert, M.~Gross,
  and B.~Hamann, Eds., Los Alamitos, 1999, pp. 19--26.

\bibitem{telea99}
A.~Telea and J.~J. van Wijk, ``Simplified representation of vector fields,'' in
  \emph{Proc.\ IEEE Visualization '99}, D.~Ebert, M.~Gross, and B.~Hamann,
  Eds., Los Alamitos, 1999, pp. 35--42.

\bibitem{SegmentationPDEs}
K.~Padberg-Gehle, S.~Reuther, S.~Praetorius, and A.~Voigt, ``Transfer
  operator-based extraction of coherent features on surfaces,'' in
  \emph{Topological Methods in Data Analysis and Visualization IV}, H.~Carr,
  C.~Garth, and T.~Weinkauf, Eds.\hskip 1em plus 0.5em minus 0.4em\relax Cham:
  Springer International Publishing, 2017, pp. 283--297.

\bibitem{AnisotropicSegmentationPDE}
U.~{Diewald}, T.~{Preusser}, and M.~{Rumpf}, ``Anisotropic diffusion in vector
  field visualization on euclidean domains and surfaces,'' \emph{IEEE
  Transactions on Visualization and Computer Graphics}, vol.~6, no.~2, pp.
  139--149, April 2000.

\bibitem{Park:2006:SADV}
\BIBentryALTinterwordspacing
S.~W. Park, H.~Yu, I.~Hotz, O.~Kreylos, L.~Linsen, and B.~Hamann,
  ``Structure-accentuating dense flow visualization.'' in \emph{EuroVis}, B.~S.
  Santos, T.~Ertl, and K.~I. Joy, Eds.\hskip 1em plus 0.5em minus 0.4em\relax
  Eurographics Association, 2006, pp. 163--170. [Online]. Available:
  \url{http://dblp.uni-trier.de/db/conf/vissym/eurovis2006.html#ParkYHKLH06}
\BIBentrySTDinterwordspacing

\bibitem{probHollt}
T.~Höllt, M.~Hadwiger, O.~Knio, and I.~Hoteit, ``Probability maps for the
  visualization of assimilation ensemble flow data,'' 01 2015.

\bibitem{Guo}
H.~Guo, W.~He, S.~Seo, H.-W. Shen, E.~Constantinescu, C.~Liu, and T.~Peterka,
  ``Extreme-scale stochastic particle tracing for uncertain unsteady flow
  visualization and analysis,'' \emph{IEEE Transactions on Visualization and
  Computer Graphics}, vol.~25, pp. 1--1, 07 2018.

\bibitem{Otto}
M.~Otto, T.~Germer, H.-C. Hege, and H.~Theisel, ``Uncertain 2d vector field
  topology,'' \emph{Comput. Graph. Forum}, vol.~29, pp. 347--356, 05 2010.

\bibitem{Matvienko:2012:DFVU}
\BIBentryALTinterwordspacing
V.~Matvienko and J.~Kr{\"u}ger, ``Dense flow visualization using wave
  interference,'' in \emph{IEEE Pacific Visusalization}, 2012, pp. 129--136.
  [Online]. Available:
  \url{http://www.ivda.uni-saarland.de/fileadmin/publications/2012/pvis2012.pdf}
\BIBentrySTDinterwordspacing

\bibitem{Shi:1997:NCIS}
J.~Shi and J.~Malik, ``Normalized cuts and image segmentation,'' \emph{IEEE
  Transactions on Pattern Analysis and Machine Intelligence}, vol.~22, pp.
  888--905, 1997.

\bibitem{Meila:2001:LSRW}
M.~Meila and J.~Shi, ``Learning segmentation by random walks,'' in \emph{In
  Advances in Neural Information Processing Systems}.\hskip 1em plus 0.5em
  minus 0.4em\relax MIT Press, 2001, pp. 873--879.

\bibitem{LaplacianEigenvalueFrequency}
D.~I. Shuman, S.~K. Narang, P.~Frossard, A.~Ortega, and P.~Vandergheynst, ``The
  emerging field of signal processing on graphs: Extending high-dimensional
  data analysis to networks and other irregular domains,'' 2012.

\bibitem{wegenkittl97}
R.~Wegenkittl, H.~L{\"o}ffelmann, and E.~Gr{\"o}ller, ``Fast oriented line
  integral convolution for vector field visualization via the internet,'' in
  \emph{Proc.\ IEEE Visualization '97}, R.~Yagel and H.~Hagen, Eds., 1997, pp.
  309--316.

\bibitem{SPARSPAK}
\BIBentryALTinterwordspacing
A.~George and E.~Ng, ``A new release of sparspak: The waterloo sparse matrix
  package,'' \emph{SIGNUM Newsl.}, vol.~19, no.~4, p. 9–13, Oct. 1984.
  [Online]. Available: \url{https://doi.org/10.1145/1057931.1057933}
\BIBentrySTDinterwordspacing

\bibitem{YalePackage}
\BIBentryALTinterwordspacing
S.~C. Eisenstat, H.~C. Elman, M.~H. Schultz, and A.~H. Sherman, ``The (new)
  yale sparse matrix package,'' in \emph{Elliptic Problem Solvers}, G.~Birkhoff
  and A.~Schoenstadt, Eds.\hskip 1em plus 0.5em minus 0.4em\relax Academic
  Press, 1984, pp. 45 -- 52. [Online]. Available:
  \url{http://www.sciencedirect.com/science/article/pii/B9780121005603500093}
\BIBentrySTDinterwordspacing

\bibitem{LaplacianEigenvectorAmplitude}
D.~S. Grebenkov and B.-T. Nguyen, ``Geometrical structure of laplacian
  eigenfunctions,'' 2012.

\bibitem{ARPACK}
\BIBentryALTinterwordspacing
\emph{Introduction to ARPACK}, pp. 1--7. [Online]. Available:
  \url{https://epubs.siam.org/doi/abs/10.1137/1.9780898719628.ch1}
\BIBentrySTDinterwordspacing

\bibitem{PETSc}
S.~Abhyankar, J.~Brown, E.~M. Constantinescu, D.~Ghosh, B.~F. Smith, and
  H.~Zhang, ``Petsc/ts: A modern scalable ode/dae solver library,'' \emph{arXiv
  preprint arXiv:1806.01437}, 2018.

\bibitem{Eigen}
\BIBentryALTinterwordspacing
G.~Guennebaud, B.~Jacob \emph{et~al.}, ``Eigen v3,'' 2010. [Online]. Available:
  \url{http://eigen.tuxfamily.org}
\BIBentrySTDinterwordspacing

\bibitem{KruegerDataset}
J.~Kr{\"u}ger, T.~Schiwietz, P.~Kipfer, and R.~Westermann, ``Numerical
  simulations on pc graphics hardware,'' in \emph{Recent Advances in Parallel
  Virtual Machine and Message Passing Interface}, D.~Kranzlm{\"u}ller,
  P.~Kacsuk, and J.~Dongarra, Eds.\hskip 1em plus 0.5em minus 0.4em\relax
  Berlin, Heidelberg: Springer Berlin Heidelberg, 2004, pp. 442--449.

\bibitem{gerrisflowsolver}
\BIBentryALTinterwordspacing
S.~Popinet, ``Free computational fluid dynamics,'' \emph{ClusterWorld}, vol.~2,
  no.~6, June 2004. [Online]. Available: \url{http://gfs.sf.net/}
\BIBentrySTDinterwordspacing

\bibitem{Candelaresi2011a}
S.~Candelaresi and A.~Brandenburg, ``Decay of helical and nonhelical magnetic
  knots,'' \emph{Phys. Rev. E}, vol.~84, no.~1, pp. 16\,406--16\,416, 2011.

\bibitem{numSim1}
\BIBentryALTinterwordspacing
M.~Griebel, T.~Dornseifer, and T.~Neunhoeffer, \emph{Numerical Simulation in
  Fluid Dynamics}.\hskip 1em plus 0.5em minus 0.4em\relax Society for
  Industrial and Applied Mathematics, 1998. [Online]. Available:
  \url{https://epubs.siam.org/doi/abs/10.1137/1.9780898719703}
\BIBentrySTDinterwordspacing

\bibitem{numSim2}
R.~Croce, M.~Griebel, and M.~Schweitzer, ``Numerical simulation of bubble and
  droplet deformation by a level set approach with surface tension,''
  \emph{International Journal for Numerical Methods in Fluids}, vol.~62, pp.
  963 -- 993, 01 2009.

\bibitem{ABC}
G.~Haller, ``An objective definition of a vortex,'' \emph{Journal of Fluid
  Mechanics}, vol. 525, pp. 1 -- 26, 02 2005.

\bibitem{Knyazev:2001:TOP:587151.587187}
\BIBentryALTinterwordspacing
A.~V. Knyazev, ``Toward the optimal preconditioned eigensolver: Locally optimal
  block preconditioned conjugate gradient method,'' \emph{SIAM J. Sci.
  Comput.}, vol.~23, no.~2, pp. 517--541, Feb. 2001. [Online]. Available:
  \url{http://dx.doi.org/10.1137/S1064827500366124}
\BIBentrySTDinterwordspacing

\bibitem{Haller:2001:DMCS}
G.~Haller, ``Distinguished material surfaces and coherent structures in
  three-dimensional fluid flows,'' \emph{Phys. D}, pp. 248--277, 2001.

\end{thebibliography}

%

\begin{IEEEbiography}[{\includegraphics[width=1in,height=1.25in,clip,keepaspectratio]{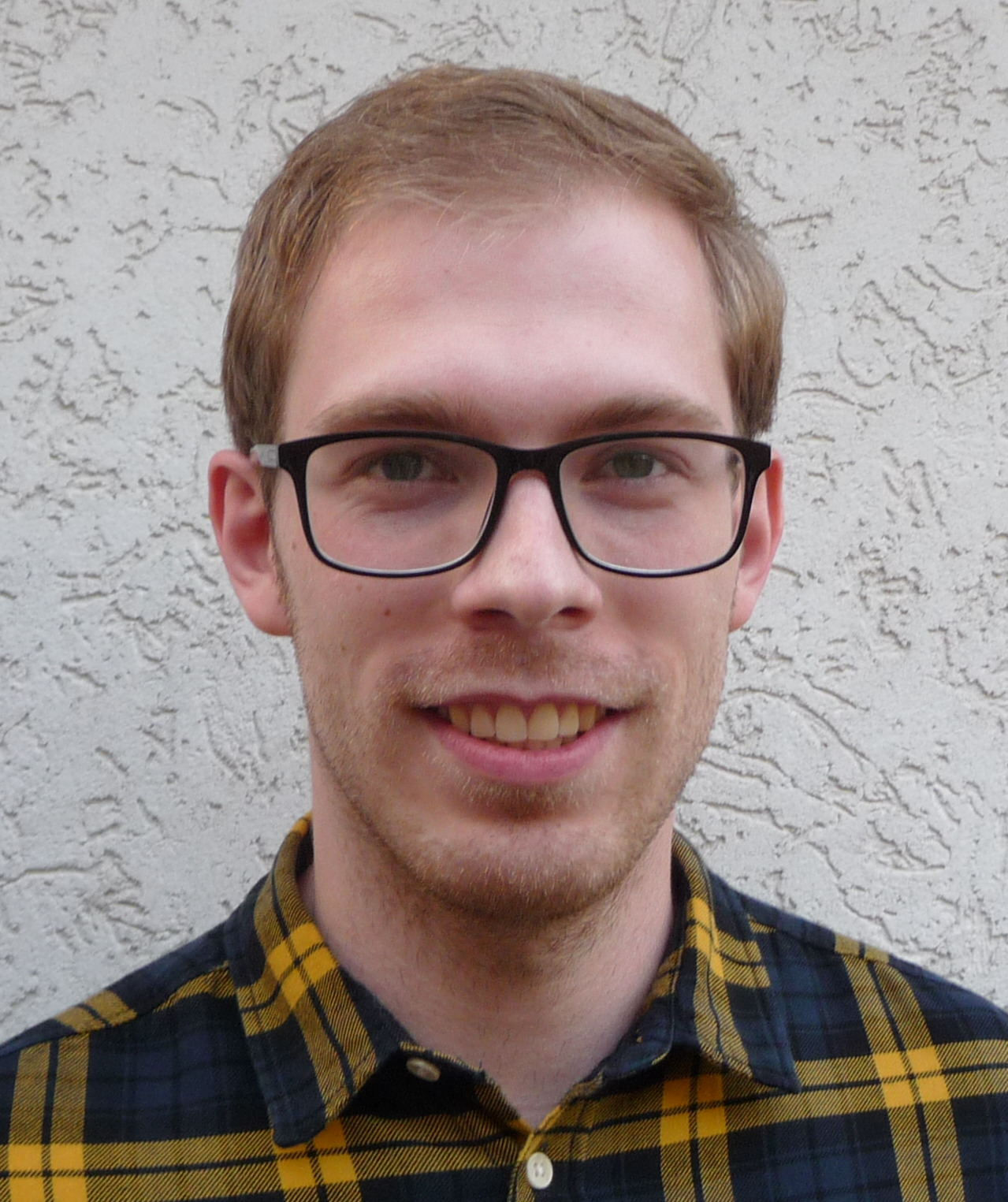}}]{Daniel Preu\ss{}}
studied technomathematics with computer science as an elective at the University of Duisburg-Essen. He received a Master's Degree in 2018. Since then, he is a PhD student at the High Performance Computing group at the University of Duisburg-Essen. His main research interests are flow visualization and analysis.
\end{IEEEbiography}

\begin{IEEEbiography}[{\includegraphics[width=1in,height=1.25in,clip,keepaspectratio]{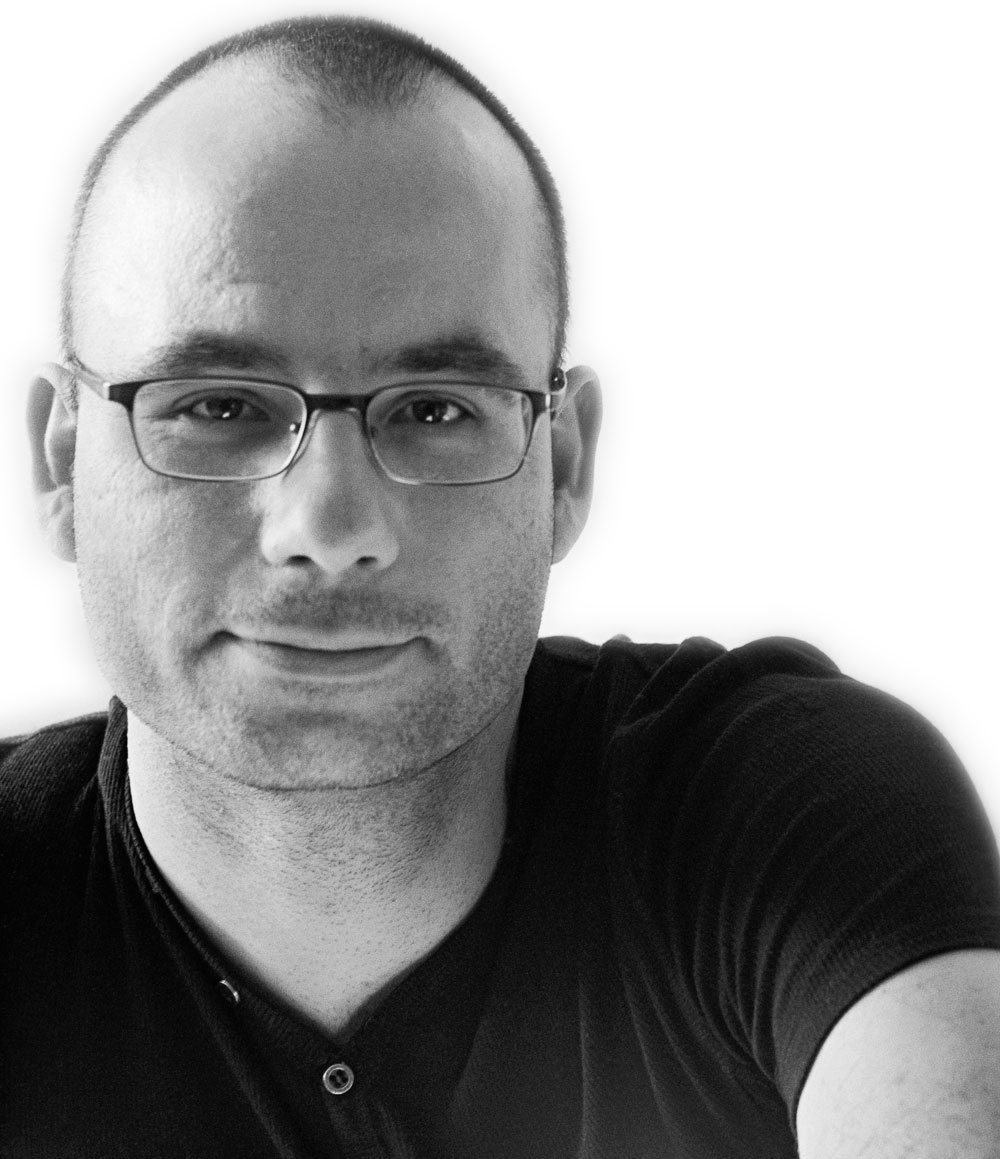}}]{Tino Weinkauf}
received his diploma in computer science from the University of Rostock in 2000.
From 2001, he worked on feature-based flow visualization and topological data analysis at Zuse Institute Berlin.
He received his Ph.D. in computer science from the University of Magdeburg in 2008.
In 2009 and 2010, he worked as a postdoc and adjunct assistant professor
at the Courant Institute of Mathematical Sciences at New York University.
He started his own group in 2011 on Feature-Based Data Analysis 
in the Max Planck Center for Visual Computing and Communication, Saarbrücken.
Since 2015, he holds the Chair of Visualization at KTH Stockholm. 
His current research interests focus on flow analysis, 
discrete topological methods, and information visualization.
\end{IEEEbiography}


\begin{IEEEbiography}[{\includegraphics[width=1in,height=1.25in,clip,keepaspectratio]{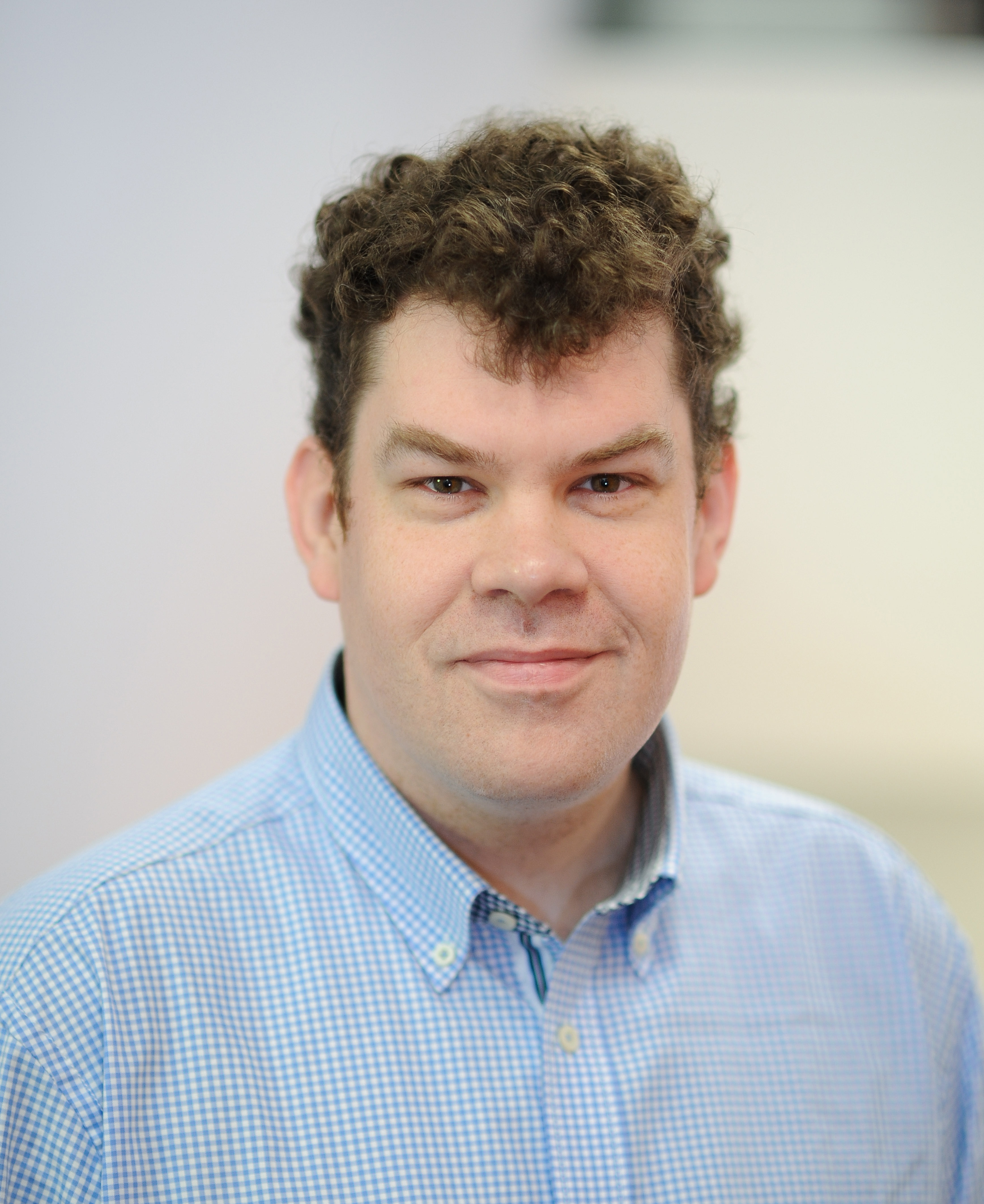}}]{Jens Kr\"{u}ger}
studied computer science at the Rheinisch-Westf\"{a}lische Technische Hochschule Aachen 
where he received his diploma in 2002. 
In 2006 he finished his PhD at the Technische Universität München and 
after Post Doc positions in Munich and at the Scientific Computing and Imaging (SCI) Institute 
he became research assistant Professor at the University of Utah.
In 2009 he joined the Cluster of Excellence Multimodal Computing and Interaction at Saarland University 
to head the Interactive Visualization and Data Analysis group.
Since 2013, Jens Kr\"{u}ger has been Chair of the High Performance Computing group at the University of Duisburg-Essen.
In addition to his position in Duisburg-Essen, 
he also holds an adjunct professor title of the University of Utah 
and is a principal investigator of multiple projects in the Intel Visual Computing Institute.
\end{IEEEbiography}




\end{document}